\documentclass[prb,apsrev,preprint,floatfix]{revtex4-1}% macair
\usepackage{amsmath}
\usepackage{graphicx,color}
\usepackage{amsfonts}
\usepackage{amssymb}
\usepackage{graphicx}%
\providecommand{\U}[1]{\protect\rule{.1in}{.1in}}

\begin{document}

\title{
\color{black}
{A quantum dynamics method for}
\color{black}
excited electrons in 
\color{black}
molecular aggregate system 
\color{black}
using a group diabatic Fock matrix}

\date{\today}
\author{Takehiro Yonehara}
\email{takehiro.yonehara@riken.jp}
\author{Takehito Nakajima}
\email{nakajima@riken.jp}
\affiliation
{RIKEN Advanced Institute for Computational Science, Kobe 650-0047, Japan}

\begin{abstract}
\baselineskip 7mm 
We introduce a practical calculation scheme 
for the description of excited electron dynamics in 
\color{black}
molecular aggregate systems 
\color{black}
within 
\color{black}
a local group diabatic Fock representation. 
\color{black}
This scheme makes it easy to analyze the interacting time-dependent excitation 
of local sites in complex systems. 
In addition, light-electron couplings are considered. 
The present scheme is intended for investigations on 
the migration dynamics of excited electrons 
in light-induced energy transfer systems.
The scheme was applied to two systems: 
a naphthalene(NPTL)-tetracyanoethylene(TCNE) dimer
and a 20-mer circle of ethylene molecules.  
Through local group analyses of the dynamical electrons,  
we obtained an intuitive understanding of the electron transfers 
between the monomers.  

\end{abstract}

\baselineskip 6mm

\maketitle

% \ tableofcontents

\baselineskip 7mm

% \ clearpage 

\section{Introduction}
\label{sect-intro}

\color{black}
The interaction of light with molecular systems is a fundamental factor 
for energy source in nature.
The photo chemistry as a representative process involved with this type of interaction 
is characterized by excited electrons in molecular systems. 
\color{black}
Especially, the dynamics of excited electrons in aggregated molecular systems 
play a significant role in energy transfer phenomena. 
\cite{Kuhn-CET,Scholes-rev-excitons,Schiffer-Rev-PCET,Spano-EET}
%
% repl.modi.7/11/2017  (  )
% discrimination of charge and exciton transfer
%
\color{black}
Electron, radical and exciton transfers are important topic 
in a research field treating excited electron dynamics in molecular aggregates. 
The electron transfer is characterized by inter or intra molecular charge migration. 
The radical electron is well described by unpaired singly occupied electrons 
in a natural orbital representation. The exciton transfer is caused 
by interactions of excitons created in coherently coupled monomer group 
as a hole-electron pair often accompanied with their delocalization.  
\color{black}
%
% ^ ^ ^ 
% Exciton delocalization, charge transfer, and electronic coupling
% for singlet excitation energy transfer between stacked nucleobases
% in DNA: An MS-CASPT2 study
% L. Blancafort and A. A. Voityuk
% J. Chem. Phys. 140, 095102 (2014)
%
%\color{black}
%We note that an understanding of the quantum dynamics of excited electrons 
%will also support an exploration of efficient energy conversion systems. 
%\color{black}
From the viewpoint of real time dynamics in a light energy conversion,  
several pioneering investigations have been reported 
in the fields of photo synthesis, solar cells, and photo chemical reactions. 
\cite{art-ps,rt-exp-pcbm,rt-exp-Graetzel,env-assist-qw-eet}

Theories for the treatment of real-time electron dynamics 
in complex excited states accompanied by the changes 
in chemical bonding, have been developed and applied, 
for example, real-time time dependent density functional theory, 
time dependent configuration interaction, and configuration state function schemes. 
\cite{RG-TDDFT,octpus-rt-tddft,gceed,PCCP-eldt-KT-TY,Nonadiabaticity-ChemRev,Chemical-theory-beyond-BO-paradigm} 
These theories provide an excellent description of electron dynamics in total systems.
However, there is no explicit scheme 
for the analysis of the time-dependent interaction dynamics 
of excited electrons in a complex system with using 
a reliable level of ab initio electronic structure calculation. 
The schemes described are successful but time-consuming, 
so a general theoretical tool allowing 
the compact description and investigation of 
the electron dynamics in any aggregated molecular would be useful.

In the present work, we propose a practical theory for describing 
the excited electron dynamics in  aggregated molecular systems 
\color{black}
driven by the interactions between subsystems and by light-matter interactions. 
\color{black}

This treatment enables us to treat arbitrary types of 
initial local excitation in a complex system.  
We have used the well-known group diabatic matrix representation of Fock operator. 
\cite{Lan,Gao,Shi,Thoss-TiO2} 
We call this the GDF model. 
A skeleton Fock matrix of the total system is represented 
by a set of localized orbitals of the sub groups 
after applying a L\"{o}wdin orthogonalizaion of the atomic orbitals.
\cite{Szabo}
In this article, we show that this provides a useful scheme for studying 
the electron dynamics accompanied by internally/externally induced excitation. 
Combined with light-matter couplings and the skeleton Fock matrix mentioned above, 
we can extract the essential features of the electron migration dynamics 
at an early stage after any type of excitation. 
Thus, our scheme can, potentially, provide meaningful information 
for the study of the transfer of light-energy to chemical functionality 
via molecular aggregate systems.
We show the efficiency of the introduced theoretical method 
by two example applications, 
a typical donor-acceptor dimer system and 
\color{black}
an aggregate of circularly arranged monomers
\color{black}
under external continuum radiation fields.
The features found in the dynamics of the excited electrons 
in these aggregated systems validate the efficiency of the present scheme.

The contents of this report are as follows: 
In Sect. \ref{sect-theory}, 
we review current theoretical methods concerning electron dynamics 
using the equations of motions 
and explain the need to introduce a group interaction representation 
to the electron dynamics theory. 
Then we explain the details of the construction of the GDF representation. 
Further details of the equation of motion of the density matrix in this representation 
are also given in this section. 
In Sect. \ref{sect-app}, we present numerical applications of the 
present method by using two types of aggregated systems.  
In the last section \ref{sect-summary}, we provide concluding remarks.

%\clearpage 

\section{Theoretical method}
\label{sect-theory}

\subsection{Computational theories for the study electron dynamics}

Here, we review the theories of electron dynamics in molecular systems 
and we explain the need for our newly devised theoretical method in this article. 
Although the interplay between the nuclear and electronic wave packets is 
an important topic, 
\cite{PSANB-laser,en-coins-JPCA,Bohmian-Suzuki-2016,Factorization-EN-Gross}
we have not discussed this to allow us to focus on 
our electron dynamics description scheme.

There are several types of practical theories 
for the description of  electron dynamics involving  excited states in molecular systems. 
\color{black}
These theories are based on the electronic structure calculations.
\color{black}
For brevity, they can be roughly categorized into two schemes: 
those directly utilizing 
(a)  a time dependent multi-electron wavefunction 
\color{black}
     as a time-dependent linear combination of configuration state function 
\color{black}
     constructed from a molecular orbital set and 
(b)  a time-dependent electron density matrix made from
     independent orbitals propagating at the electron timescale.

First, in the following two subsections, 
we review separately these two schemes, and  in the last subsection, 
we address the differences with a related scheme 
for the treatment of the interaction dynamics in excited aggregated systems.

\subsubsection{\textrm{(a)} Schemes based on multi-state multi-electron wavefunctions}
%
%%%
%%% (a)
%%%
The time-dependent configuration interaction theory (TDCI)  
including the time dependent configuration state function (TDCSF) 
\cite{Saalfrank-JCP2007-TDCI,TDCSF-Amano,Nonadiabaticity-ChemRev,Chemical-theory-beyond-BO-paradigm}
and multi configurational time-dependent Hartree-Fock (MCTDHF) schemes
\cite{mctdhf-scrinzi,mctdhf-kato-kono}
are representatives of the class (a).

In the TDCI scheme, 
the propagation of the electron wave packet obeys 
a snapshot equation of motion at each molecular configuration.
Here, an electron wave packet is defined as 
the superposition of multi-electronic adiabatic state functions of a molecular system. 
Only the configuration interaction (CI) coefficients are propagated in the time domain
while the electron orbitals for constructing the configuration basis functions 
are determined in a static manner at each molecular geometry. 
In cases considering the molecular structure dynamics,  
%the orbitals, as well as CSF uniformity with changes in the molecular geometry,
\color{black}
the orbitals uniformity as well as resultant uniformity of CSF made from orbitals  
with changes in the molecular geometry,
\color{black}
are needed 
for consistency between the time-dependent CSFs and their coefficients with dynamics. 
This includes the solution responsible for 
the orbital character exchange with molecular motion. 
The type depends on the variation of the methods 
for the compact, smooth, and  correct description of the electron dynamics.  
\cite{TDCSF-Amano,Nonadiabaticity-ChemRev,TDCSF-Kunisada-Ushiyama}

Unlike TDCI, where only the coefficients of the CI functions or CSFs 
propagate in the time domain, 
in the MCTDHF scheme, which is a Fermion system counterpart 
of the multi configuration time dependent Hartree (MCTDH) method, 
\cite{book-Meyer-MCTDH}
both the configuration coefficients and orbital parameters are simultaneously 
propagated according to the equation of motion derived 
from the application of the time-dependent variational principle to the quantum action. 
\color{black}
Here, time-dependent variational principle means 
the equation of motions for the parameters determining 
a time dependent quantum state is derived by 
the stationalization
of the quantum action $\int dt \langle \psi(t)|\hat{H}-i \hbar d/dt|\psi(t) \rangle$
as a functional of time dependent quantum state, $|\psi(t) \rangle$,  
under the constraint of normalization condition $ \langle \psi | \psi \rangle = 1$
with respect to the variation of $ | \psi(t) \rangle $
where $\hat{H}$ denotes Hamiltonian operator of a system under consideration.
\color{black}
Because of this substantially instantaneous optimization of the time dependent orbitals 
by using the effective one-electron self-consistent field, 
in MCTDHF-type of schemes, 
we can save the dimensions of the multi-electron configuration functions. 
This approach is usually applied to high-accuracy studies 
of field induced electron dynamics in small molecular systems with high accuracy. 
As practical schemes accompanied by 
the modelling of a time-dependent active orbital space in a sophisticated manner,  
the time-dependent complete active space self-consistent field (TDCASSCF) 
\cite{tdcascf-sato} theory
and the restricted active space counterpart (TDRASSCF) 
\cite{tdrasscf-Madsen}
were developed in a similar manner to that of MCTDHF theory.

One advantage of the schemes belonging to (a) is 
that they offer a picture of the electron wave packet dynamics
in terms of multiple multi-electron adiabatic or pseudo-diabatic states.  
For example, this aspect allows the theoretical analysis 
of nonadiabatic and optical transitions between multi electronic adiabatic states. 
The methods categorized into (a) can principally 
reap the benefit of high level electronic correlation methods 
including post-Hartree-Fock theories. 
\cite{TD-CAS-CISD-Nest,TD-DMRG}

Applications with using (a) include 
a field-induced electron dynamics at the 
\color{black}
attosecond 
\color{black}
timescale, 
\cite{Diborane-laser,PCCP-eldt-KT-TY}
nonadiabatic electron wave packets in a proton-coupled electron transfer processes, 
\cite{Ushiyama-ACIE2006-PCET,nagashima2012,TDCSF-pcet-yama-taka}
the characterization of electron dynamics in highly quasi-degenerate excited states 
\cite{b12-nad}, 
and ultrafast photoionization process.
\cite{MuCurdy-MCTDHF-appl-ionization,TDCSF-ionization-matsu-taka}
Despite the successes in analyzing electron dynamics at the 
\color{black}
attosecond 
\color{black}
scale, 
the system size is inevitably limited by the computational cost 
associated with the complex mathematical ingredients of these methods, 
thus reducing the ease of their use.

\subsubsection{\textrm{(b)} Schemes based on the electron density matrix}
%%%
%%% (b)
%%%
%
%
The class (b) includes the time dependent Hartree-Fock (TDHF) and, 
\cite{Kulander-PRA-rt-TDHF}
real-time time-dependent density functional theories (RT-TDDFT).
\cite{RG-TDDFT,octpus-rt-tddft,gceed,Meng-Kaxiras-JCP2004-rt-tddft-AO}
Here, the electron density operators and associated matrices play central roles 
in the description of real-time electron dynamics.
In particular, compared to the class (a) 
RT-TDDFT provides a way to treat large molecular systems 
by considering the electronic correlation at a moderate level 
using effective approximations, resulting  
in a reasonable computational cost.

% 
% orb.propagation
% 
There are two types for RT-TDDFT. 
One utilizes time-dependent mutually independent orthogonal orbitals 
obeying the time-dependent Kohn-Sham (TDKS) equation, 
\cite{Rubio-JCP2004-propagation-tdks}
where a time-dependent one electron field is 
determined by the electron density matrix constructed via 
the ground state occupation of the TDKS orbitals 
in the same way as that of density functional theory (DFT). 
Note that the determination of the occupied orbitals is equivalent to 
that of the one-electron density matrix.
In this orbital scheme, 
all the occupied orbitals are directly propagated at the electron timescale.  
The equation of motion of the $i$-th TDKS orbital, $\phi_i(t)$, 
is written as 
\begin{align}
\dfrac{d}{dt}\phi_i(t) 
= - \dfrac{i}{\hbar} \hat{H_0}[\{\phi_{\cdot}\}] \phi_i(t), 
\label{KS-propagator-form}
\end{align}
where $i$ is an orbital label and runs over all occupied orbitals.  
The Kohn-Sham Fock operator, $\hat{H}_0$, depends on 
the occupied independent Kohn-Sham orbitals. 
This is expressed by Eq. (\ref{KS-propagator-form}) 
as a set of functions $\{\phi_.\}$ 
\color{black}
in the parentheses $[...]$, 
\color{black}
attached to $\hat{H}_0$. 
Here $ \{\phi_.\}$ means whole the set of occupied TDKS orbitals under consideration.

%
% den.mat.propagation
%
The other scheme of RT-TDDFT is based on 
the direct time propagation of the electron density matrix 
without any propagation of the KS orbitals. 
\cite{Voorithr-rt-TDDFT-density-mat}
The KS orbitals are determined through a self-consistent field calculation 
that is carried out once at the initial simulation time 
and constructs the initial electron density matrix 
needed for a time propagation.
Formally, the time propagation of density matrix, $\underline{\underline{\rho}}$, 
within any basis set representation  
obeys the Liouville von Neumann equation, including the one electron effective Hamiltonian, 
which is constructed as the sum of the time-dependent Fock matrix, 
$\underline{\underline{F}}$, 
and light-electron coupling Hamiltonian, $\underline{\underline{L}}$, 
% LvN eq
\begin{align}
\dfrac{\partial}{\partial t}
\underline{\underline{\rho}} 
=
-\dfrac{i}{\hbar}
\left[
\,
\underline{\underline{F}}[\underline{\underline{\rho}}]
+
\underline{\underline{L}}
, 
\,
\underline{\underline{\rho}} 
\,
\right].  
\label{VNLeq-formal}
\end{align}
Here, $[A,B] \equiv AB-BA$.
Note that the Fock matrix depends on the time-dependent density matrix 
at each electron propagation time-step. 
Therefore, the updating the constructed Fock matrix is 
the dominant part of the computation. 
Usually some approximations with respect to 
the update of the Fock matrix and propagation schemes 
in this non-linear type equation of motion are employed 
to reduce computational cost.

%
% equivalent 
%
%These two schemes using orbitals and densities are equivalent 
%within the framework of TDDFT framework  
%under the adiabatic approximation for electron functionals named 
%as time-in-local one and orbital occupation manner 
%for the time dependent KS orbitals as referential independent orbitals.  
\color{black}
These two schemes using orbitals and densities are equivalent to each other 
within the TDDFT framework under the adiabatic approximation for electron functionals named as 
time-in-local one and orbital occupation manner for the time dependent KS orbitals 
as referential independent orbitals.
\cite{book-fundamental-TDDFT-Gross}
\color{black}
In the main part of this article, 
we use the density propagation scheme as explained 
in this subsection but with an extra approximation.

\subsubsection{Group analysis of electron dynamics in an aggregated system}

In previous studies, a time dependent interaction analysis was not sufficient 
to investigate the excited electron migration dynamics 
in molecular aggregate systems. 
Below, we describe two theoretical works related to our approach and our aims. 

 The first approach uses
 the molecular orbital (MO)-CI Quantum Master Equation (MOQME) method 
 \cite{MOQME} developed by Kishi and Nakano. 
 This theory can treat the electron dynamics in an excited molecular aggregate 
 by using a quantum master equation and density matrix within a CI representation.  
 This has been successfully applied to donor-acceptor systems and has 
 revealed the mechanism of exciton recurrence motions that are dependent on 
 an external laser fields. 
 The differences between this method and ours  are threefold. 
 The first difference concerns the type of ab initio calculation.    
 Kishi and Nakano used density matrices associated with the CIS 
 multi electron wavefunctions. 
 In contrast, our scheme is a Kohn Sham Fock based method that includes 
 effective dynamic correlation.
 The second difference is that Kishi and Nakano used 
 the Born-Markov approximation based on the weak interaction of 
 monomers in the aggregated system; we did not use this type of approximation 
 to avoid the inherent loss of information, which would limit the future extension of our method. 
 The third difference is that 
\color{black}
 our scheme can use any group separation for group local diabatization 
 in the representation of electron property matrices, including the density matrix, 
 which allows the explicit analysis of group interactions in a time-dependent manner. 
\color{black}

\color{black}
 The second approaches are the study of electron transfer dynamics 
 in a test molecular donor-acceptor pair for use as a solar material 
 and a realistic model of photo energy conversion material by using 
 the GDF Hamiltonian and the advanced version of MCTDH method (multi-layer MCTDH) 
 by Thoss \cite{Thoss-TiO2} and Xie {\it et al.} \cite{Lan}, respectively. 
 The former author, Thoss, is one of the founders of the 
 multi-layer MCTDH \cite{ML-MCTDH} scheme 
 which made breakthrough in a higher dimensional quantum dynamics study 
 and also made intensive researches by using this block diagonal approach. 

 Thoss {\it et al.} examined the photo-induced electron-transfer process 
 in the alizarin-TiO2 system as a dye-semiconductor system by using the vibronic model Hamiltonian 
 obtained by the first principle calculation and the block diagonalization approach. 
 They found an interesting feature that an electron injection process 
 proceed in a femtosecond time scale, which is assisted 
 by a significant electronic coherence.
\color{black}

 In the work by Xie {\it et al.}, they constructed a four-state electronic Hamiltonian
 using a GDF representation and DFT calculations and then  
 investigated the multi-electronic-state nuclear wave packet dynamics 
 using 4 electronic states and 246 vibrational modes.
 They carefully chose the electronic states responsible for 
 the charge transfer dynamics and, 
 by increasing the number of vibrational modes, 
 obtained valuable information 
 concerning the effect of the vibrational motion on the electron dynamics.
 Although their work is excellent and 
 partly analyzed the electronic properties as diabatic state populations 
 by using the nuclear wave packet dynamics,
 it remains challenging to 
 examine the associated electronic properties based on a localized representation, 
 as well as, for example, unpaired electrons and bond order properties, among others. 
 For example, in the cases with external laser fields, 
 many more electronic states should be considered. 
 However, this is difficult to treat using the approaches mentioned above.

\color{black}
A transparent as well as convenient group analysis scheme 
for describing electron dynamics is required, not only for the investigation 
of the unresolved excited electron affinities between or within constituent molecules.  
\cite{Scholes-rev-excitons}

In the following parts of this article, 
a description of our theoretical methodology for this purpose 
is given, and its validity is verified through application 
to two types of aggregated systems.  
\color{black}

\subsection{Group diabatic representation}
\label{sect-GDF}

On constructing a group diabatic representation of the effective electronic Hamiltonian 
for the examination of the excited electron dynamics in a general molecular aggregated system, 
we employed the procedures applied in the articles of Lan, Gao, Shi and Thoss.
\cite{Lan,Gao,Shi,Thoss-TiO2}
However, here, we have added a density matrix time propagation scheme 
and light-electron couplings to their formulation. 
Below we summarize this briefly.

\subsubsection{Fock matrix with orthogonalized L\"{o}wdin atomic orbitals}
We employ the representation matrix of the Fock operator 
in terms of the well-known L\"{o}wdin orthogonalized atomic basis function, 
\cite{Szabo}
\begin{align}
\widetilde{F}_{mn} 
\equiv
\langle \widetilde{\chi}_m | \widehat{F} | \widetilde{\chi}_n \rangle   
=
[ \underline{\underline{\widetilde{F}}} ]_{mn}
=
[
\underline{\underline{S}}^{-1/2}
\,
\underline{\underline{F}}  
\,
\underline{\underline{S}}^{-1/2}
]_{mn}, 
\end{align}
where the orthogonalized L\"{o}wdin atomic orbitals(AO) are expressed by 
\begin{align}
  | \widetilde{\chi}_n \rangle
 = \sum_j^{\textrm{AO}} | \chi_j \rangle (\underline{\underline{S}}^{-1/2})_{jn}
\end{align}
with $S_{jn} = \langle \chi_j | \chi_n \rangle$ 
being the AO overlap matrix element 
and $\{ \chi_n \} $ being the original AOs. 
The original AO representation matrix of the Fock operator, 
$ \underline{\underline{{F}}} $, is responsible for the electron density 
of arbitrary electronic states.  
\color{black}
In this study, for brevity, 
in order to focus on a minimal description of the electron dynamics, 
we used the ground state electron density at an initial simulation time 
to construct a reference Hamiltonian for the electrons 
as the first stage of introducing our proposed scheme.  
During a simulation, Fock matrix was unchanged.  
\color{black}

\subsubsection{Group localized orbitals of subgroup}

 First, we classify $\{  \widetilde{\chi}_n  \}$ into subgroups, for example, monomers. 
Here, we can safely expect that 
\color{black}
the atom center of the {\it i}-th L\"{o}wdin atomic orbital, $ \widetilde{\chi}_i $, 
obtained by L\"{o}wdin orthogonalization is 
the same as that of the original i-th AO orbital. 
\color{black}
In addition, 
arbitrary divisions of the component atoms among the total system are possible. 

 The block structure of the Fock matrix within the L\"{o}wdin basis set is 
expressed by  
\begin{align}
\underline{\underline{\widetilde{F}}}
&
=
\left(
\begin{array}{cccc}
\underline{\underline{\widetilde{F}}}_{G_1 G_1} &\underline{\underline{\widetilde{F}}}_{G_1 G_2} &.. &\underline{\underline{\widetilde{F}}}_{G_1 G_{N_g}} \\
\underline{\underline{\widetilde{F}}}_{G_2 G_1} &\underline{\underline{\widetilde{F}}}_{G_2 G_2} &.. &\underline{\underline{\widetilde{F}}}_{G_2 G_{N_g}} \\
                                ..  &..                                  &.. &.. \\ 
\underline{\underline{\widetilde{F}}}_{G_{N_g} G_1} &\underline{\underline{\widetilde{F}}}_{G_{N_g} G_2} &.. &\underline{\underline{\widetilde{F}}}_{G_{N_g} G_{N_g}} 
\end{array}
\right),
\end{align}
where we used the property of  $S$ as a real-symmetric matrix.
Here, $G_i$ denotes the i-th sub group while $N_g$ is 
the number of sub groups in the system.  

The diagonalization of diagonal block corresponding to subgroup $G$, 
\begin{align}
 \underline{\underline{\widetilde{F}}}_{GG}  
=
 \underline{\underline{D}}_{G}  
\,
 \underline{\underline{\overline{F}}}_{GG}  
\,
 \underline{\underline{D}}_{G}^{\dagger}  
\,\,
\Leftrightarrow
\,\,
 \underline{\underline{\overline{F}}}_{GG}  
=
 \underline{\underline{D}}_{G}^{\dagger}  
\,
 \underline{\underline{\widetilde{F}}}_{GG}  
\,
 \underline{\underline{D}}_{G}  
=
\left( 
\begin{array}{cccc}
\epsilon_{1,G}  & 0              &..   &  0   \\
  0             & \epsilon_{2,G} &..   &  0   \\
..              &..              &..   &..    \\
  0             & 0              &..   & \epsilon_{M_G,G} 
\end{array}
\right), 
\end{align}
gives rise to the unitary transformation matrix, 
$\underline{\underline{D}}_{G}$, 
of which the column vectors are the linear coefficient vectors of 
the localized eigenstates expanded in terms of 
the L\"{owdin} orthogonalized atomic basis functions for the group G. 
The dagger symbol attached to a matrix indicates its adjoint form. 
$ \underline{\underline{\overline{F}}}_{GG} $ is the diagonal matrix 
having eigen energies, $\{\epsilon_{j,{G}}\}_{j = 1 \sim M_{G} }$,
of the corresponding subgroup G as their elements, and $M_{G}$ 
is the number of local basis sets spanned at the group G.    
Here, $ {G} \in \{ {G}_1,...,{G}_{N_g} \} $. 
The collection of these localized group eigenstates over all groups,
$ \left\{ \underline{\underline{D}}_{G}  \right\}_{{G}={G}_1 \sim {G}_{N_g}} $, 
spans the same Hilbert space of that of the original atomic orbitals.

In the equations above we implicitly indicated the complex Hermitian forms of the 
Fock operators and related matrices because, generally, they are complex and Hermitian. 
This is the case concerning spin-orbit couplings.
However, because we do not treat spin-orbit couplings here, 
the Fock matrices in the AO and L\"{o}wdin orthogonalized AO basis representations 
are real and symmetric. Correspondingly, the transformation matrices for 
the construction of the group localized orbitals are real, orthogonal matrices.

The important feature is that these sets of localized orbitals are not 
orthogonalized between different groups;  
this provides a diabatic character in the representation with 
use of the collection of these orbital sets. 
Thus, we can obtain group localized orbital sets required for 
the GDF representation, as explained in the next subsection.

\subsubsection{Group diabatic Fock matrix}

The group diabatic representation of the Fock operator is constructed as follows.  
The divided blocks in the L\"{o}wdin representation 
$ \underline{\underline{\widetilde{F}}} $
are transformed to the group interaction representation using 
$ \left\{ \underline{\underline{D}}_G  \right\}_{G=G_1 \sim G_{N_g}} $, 
\begin{align}
 \underline{\underline{\overline{F}}}_{G_i\,G_j}  
=
 \underline{\underline{D}}_{G_i}^{\dagger}  
\,
 \underline{\underline{\widetilde{F}}}_{G_i\,G_j}  
\,
 \underline{\underline{D}}_{G_j}  
\end{align}
which yields the GDF matrix,  
\begin{align}
 \underline{\underline{\overline{F}}}
&
=
\left(
\begin{array}{cccc}
\underline{\underline{\overline{F}}}_{G_1 G_1} &\underline{\underline{\overline{F}}}_{G_1 G_2} &.. &\underline{\underline{\overline{F}}}_{G_1 G_{N_g}} \\
\underline{\underline{\overline{F}}}_{G_2 G_1} &\underline{\underline{\overline{F}}}_{G_2 G_2} &.. &\underline{\underline{\overline{F}}}_{G_2 G_{N_g}} \\
                                ..  &..                                  &.. &.. \\ 
\underline{\underline{\overline{F}}}_{G_{N_g} G_1} &\underline{\underline{\overline{F}}}_{G_{N_g} G_2} &.. &\underline{\underline{\overline{F}}}_{G_{N_g} G_{N_g}} 
\end{array}
\right).  
\end{align}
In this form, the sub-matrices in the diagonal blocks, 
$\underline{\underline{\overline{F}}}_{G_i G_i}$ $(i = 1 \sim N_g)$,
are the diagonal matrices corresponding to the local group eigen energies, 
while the sub-matrices placed at off-diagonal blocks, 
$\underline{\underline{\overline{F}}}_{G_i G_j}$ $(i \ne j)$,
describe the interactions with different blocks.

\color{black}
Note also that the Fock matrix in group diabatic representation 
does not lose any part of that in AO presentation, namely, 
no approximation is introduced during the transformation 
between these two representations.   
\color{black}

\subsubsection{Transformation of observable, Fock and density matrices}
A matrix representation of any observable operator $\hat{O}$
in group diabatic basis set, $\underline{\underline{O}}^{\mathrm{GD}}$, 
is related to that of the original AO basis set, 
$\underline{\underline{O}}^{\textrm{AO}}$, as follows: 
\begin{align}
\underline{\underline{O}}^{\mathrm{GD}} 
&
=
\underline{\underline{U}}^{\dagger} \,
\underline{\underline{O}}^{\textrm{AO}}      \,
\underline{\underline{U}} 
\label{trans-obs}
\end{align}
where 
\begin{align}
\underline{\underline{U}} \equiv 
\underline{\underline{S}}^{-1/2} \underline{\underline{W}} 
\quad
\textrm{and}
\quad
\underline{\underline{W}}
\equiv
\left(
\begin{array}{ccc}
\underline{\underline{D}}_{G_1} &.. &\underline{\underline{0}}   \\
                            ..  &.. &..                                 \\ 
\underline{\underline{0}}       &.. &\underline{\underline{D}}_{{G}_{N_g}} 
\end{array}
\right).  
\end{align}
The Fock matrix obeys the same transformation rule 
and is obtained by setting $\hat{O}=\hat{F}$ in the above equations, where 
we know $ \underline{\underline{F}}^{\textrm{AO}} = \underline{\underline{F}} $ and 
$ \underline{\underline{F}}^{\mathrm{GD}} = \underline{\underline{\widetilde{F}}} $.

In contrast, the transformation of the density matrix 
between the original AO basis set and the group diabatic one is given by 
\begin{align}
\underline{\underline{\rho}}^\mathrm{GD} 
&
=
\underline{\underline{U}}           \,
\underline{\underline{\rho}}^\mathrm{AO}   \,
\underline{\underline{U}}^{\dagger}. 
\label{trans-dens}
\end{align}
Note that the unitarity of $\underline{\underline{W}}$ assures 
total electron conservation for this transformation,  
$ \textrm{Tr} 
\left[ \underline{\underline{\rho}}^{\mathrm{AO}} \underline{\underline{S}} \right]
= \mathrm{Tr} \left[ \underline{\underline{\rho}}^\mathrm{GD} \right] $.

\subsubsection{State couplings}

In our theoretical method, 
the essential elements required for the construction of the light-electron couplings are
$ \hat{O} = \hat{\bf r} $,  $ \partial_{\bf r} $, related to Eq. (\ref{trans-obs}).  
Here, 
boldface denotes a vector in three-dimensional Cartesian space, 
and ${\bf r}$ denotes a composite variable of the electron position in 
three-dimensional space.
The first and second operators are responsible for the light-electron couplings 
in length and velocity forms. \cite{Faisal-LG-VG,ACP-laser}
In this article, we neglect the non-adiabatic coupling and 
molecular motion to allow us to focus on the present electron dynamics scheme
using the group diabatic representation.

\subsubsection{Time propagation of density matrix in GD representation}

In the group diabatic representation explained in the previous subsection, 
the electron dynamics are naturally described in terms of 
the Liouville-von Neumann equation and the associated density matrix as follows: 
%\begin{align}
$
\dfrac{\partial}{\partial t}
\underline{\underline{\rho}}^\mathrm{GD} 
=
-\dfrac{i}{\hbar}
\left[
\underline{\underline{\mathcal{F}}}^\mathrm{GD}, 
\underline{\underline{\rho}}^\mathrm{GD} 
\right], 
\label{VNLeq}
$
%\end{align}
where 
%\begin{align}
$
\underline{\underline{\mathcal{F}}}^\mathrm{GD}  
\equiv
\underline{\underline{F}}^\mathrm{GD}  +
\underline{\underline{L}}^\mathrm{GD}  
$.
%\end{align}
Here, $ \underline{\underline{L}}^\mathrm{GD} $  
is a light-electron coupling matrix.
The light-electron couplings are described by  
$
\underline{\underline{D}}^\mathrm{GD}  
=
+ e \underline{\underline{{\bf r}}}^\mathrm{GD} {\bf E} 
$
for the length gauge and 
$
\underline{\underline{D}}^\mathrm{GD}  
=
- i \hbar \dfrac{e}{c} {\bf A}
\underline{\underline{\partial_{\bf r}}}^\mathrm{GD} 
$
for the velocity gauge.
\cite{Faisal-LG-VG,ACP-laser}
Here, ${\bf E}$ and ${\bf A}$ are 
the three-dimensional electric field vector and 
electromagnetic field vector potential, which  
generally depend on a point in a three-dimensional space, respectively.  
Because the wavelength of light treated here is sufficiently large 
for the size of the molecular system treated in this article, 
we can safely employ the long wave length approximation that 
${\bf E}$ and ${\bf A}$ do not depend on the spatial location and 
depend only on time. \cite{Faisal-LG-VG,ACP-laser}
In this study, we employed the length gauge. 
The time steps were set to 4 
\color{black}
attosecond 
\color{black}
and 
we followed the dynamics for 15 
\color{black}
femtosecond 
\color{black}
in all the calculations. 
We used the fourth-order Runge Kutta time integrator. 
\color{black}
If we directly consider the core orbitals, 
the spectral range of the GDF matrix becomes too large and 
a very small time step is required for a numerical convergence 
of total electron number as a trace of density matrix during a dynamics. 
As one way to avoid this, we first set to zero sufficiently small interactions 
between the core and other orbitals in the GDF matrix
and by setting the original core orbital energies to 
a uniform value, such as the lowest valence orbital energy. 
This treatment is substantially equivalent to a neglection of 
core orbital dynamics and interaction with the valence orbitals. 
However, this does not reduce computational cost since 
the matrix dimension does not change. 
Of cause, we can extract the small size of valence and virtual space.
In such a treatment, a computational cost will be reduced.    
But, this reduction treatment is not applied in the numerical demonstrations 
in the present article.
\color{black}
The electron matrices required for the dynamics calculation, 
namely, the Fock, dipole, and electron velocity moment matrices within the AO representation 
were obtained by using the electronic structure package, NTChem2013.
\cite{NTChem}

\subsubsection{Initial density matrix: local excitation and electron filling}
\label{init-dens}

The initial density matrices in the GDF representation are set   so that 
\color{black}
diagonal elements in each diagonal block of the corresponding monomer 
should be occupied up to the number of electrons assigned to this monomer. 
The simple example will be shown later in this subsection.
\color{black}
In a restricted case with the same spatial orbitals for different 
alpha and beta spins, 
the GDF local orbitals of each monomer are occupied by up to 
the half the number of electrons assigned to this group 
from the lowest energy orbital. 
Therefore, this initial density matrix, by construction, 
differs from that of the true ground state of the whole system 
and corresponds to the density matrix of a slightly excited state.

Based on this referential configuration in the group diabatic representation 
mentioned above, we can further introduce an excitation configuration 
measured from the referential occupations of the group diabatic localized orbitals.  
To prepare the initial densities in the cases of excess or deficient electrons in each monomer, 
we set the occupations to the corresponding number of electrons in each monomer. 

\color{black}
For example, let us consider the spin-restricted case 
with three monomers each of which has 2 electrons and 3 orbitals. 
We suppose that we want to construct the initial condition 
such that the first and third monomers are initially excited from local HOMO to LUMO. 
In this case, the setting way of initial density matrix in GDF scheme is as follows: 
[1] set reference density matrix, namely GDF ground state,
$\underline{\underline{\rho}}^{\textrm{GD:ground}}$
and   
[2] carry out local HOMO LUMO excitation and obtain the aimed GDF density, 
$\underline{\underline{\rho}}^{\textrm{GD:1,3-HL}}$
namely, 
\begin{align}
\underline{\underline{\rho}}^{\textrm{GD:ground}} \equiv 
\left(
\begin{array}{ccccccccc}
2 & 0 & 0 & 0 & 0 & 0 & 0 & 0 & 0  \\ 
0 & 0 & 0 & 0 & 0 & 0 & 0 & 0 & 0  \\ 
0 & 0 & 0 & 0 & 0 & 0 & 0 & 0 & 0  \\ 
0 & 0 & 0 & 2 & 0 & 0 & 0 & 0 & 0  \\ 
0 & 0 & 0 & 0 & 0 & 0 & 0 & 0 & 0  \\ 
0 & 0 & 0 & 0 & 0 & 0 & 0 & 0 & 0  \\ 
0 & 0 & 0 & 0 & 0 & 0 & 2 & 0 & 0  \\ 
0 & 0 & 0 & 0 & 0 & 0 & 0 & 0 & 0  \\ 
0 & 0 & 0 & 0 & 0 & 0 & 0 & 0 & 0   
\end{array}
\right)
\quad
\Longrightarrow
\quad
\underline{\underline{\rho}}^{\textrm{GD:1,3-HL}} \equiv 
\left(
\begin{array}{ccccccccc}
1 & 0 & 0 & 0 & 0 & 0 & 0 & 0 & 0  \\ 
0 & 1 & 0 & 0 & 0 & 0 & 0 & 0 & 0  \\ 
0 & 0 & 0 & 0 & 0 & 0 & 0 & 0 & 0  \\ 
0 & 0 & 0 & 2 & 0 & 0 & 0 & 0 & 0  \\ 
0 & 0 & 0 & 0 & 0 & 0 & 0 & 0 & 0  \\ 
0 & 0 & 0 & 0 & 0 & 0 & 0 & 0 & 0  \\ 
0 & 0 & 0 & 0 & 0 & 0 & 1 & 0 & 0  \\ 
0 & 0 & 0 & 0 & 0 & 0 & 0 & 1 & 0  \\ 
0 & 0 & 0 & 0 & 0 & 0 & 0 & 0 & 0   
\end{array}
\right)
\end{align}
\color{black}

For simplicity of presentations in this article, 
we employed the spin-restricted representation and the electron occupations of 
the neutral monomers at an initial simulation time. 
\color{black}
We also note that here we treat cases of overall singlet spin state in this article. 
\color{black}
The details and calculation results for the spin-unrestricted and charged cases 
will be reported in our future articles.

By using this scheme, we can analyze the dynamical electron re-distribution 
in molecular aggregate systems 
driven by the differences in the electron affinities between local groups 
and the optical-electron interactions.

\section{Numerical applications}
\label{sect-app}

We examined two types of systems as a demonstration of the present scheme.  
The first example concerns 
the charge migration dynamics triggered by an external light field 
in a naphthalene(NPTL)-tetracyanoethylene(TCNE) dimer, 
which used as a typical electron donor-acceptor system;  
the results are used as the first illustration of our scheme.  
The second example is 
the unpaired electron dynamics in an excited 20-mer ethylene system. 
This is used as a model of exciton transfer within a molecular ring system.  
%
%  revise1 7/13/2017
%
\color{black}
Though the component monomers are different,   
such a ring shape is found in one unit of light-harvesting antenna systems   
\cite{LH-anthena-ring,MOQME} 
where, an internal electron migration can affect an energy transfer among ring aggregates. 
The ring shape also found in a test system consisting of weakly interacting excited atoms 
at ultra cold temperatures as an example of exotic system 
for the experimental examination of the mechanism of quantum dynamics and its control by 
an external field associated with highly structured excited states. 
\cite{atom-ring-Ryd-spin,Eisfeld-ultracold-coins}
\color{black}
In both cases, the time propagations of the density matrix were carried out  
in the Hilbert spaces spanned by all the AOs. 
\color{black}
In the present article, for simplicity,  nonadiabatic couplings are omitted and 
cases of spin singlet state for the total systems are treated. 
\color{black}

\subsection{Charge migration dynamics: NPTL-TCNE dimer}

Here, by applying the present theoretical method to a NPTL-TCNE dimer system, 
we observed the initiation of charge separation triggered by the electron-light interaction, 
where
\color{black}
%the light had the ordinary wave length of sunlight and 
%appropriate polarization vectors along the molecular axes.
the light had the ordinary wave length (700 nm) of sunlight
and distinct polarization vectors along molecular axes.
The balance of the photon energy and the energy gaps of 
localized orbitals of monomers are discussed 
in the supplementary material with respect to the mode analysis of electron dynamics. 
\color{black}
Because of the properties with respect to electron addition, as well as ejection
(the adiabatic electron affinity of TCNE is large (3.16 eV)
\cite{NIST,adia-elec-affinity-TCNE}, as is  
the adiabatic ionization potential of NPTL (8.14 eV)
\cite{NIST,ionization-energy-NPTL}), 
the NPTL-TCNE set makes an ideal donor-acceptor molecular pair 
for checking the description of charge transfer with using the present GDF scheme.

%
% geometry configuration of dimer NPTL-TCNE
%
\begin{figure}[th]
\includegraphics[width=0.55\textwidth]{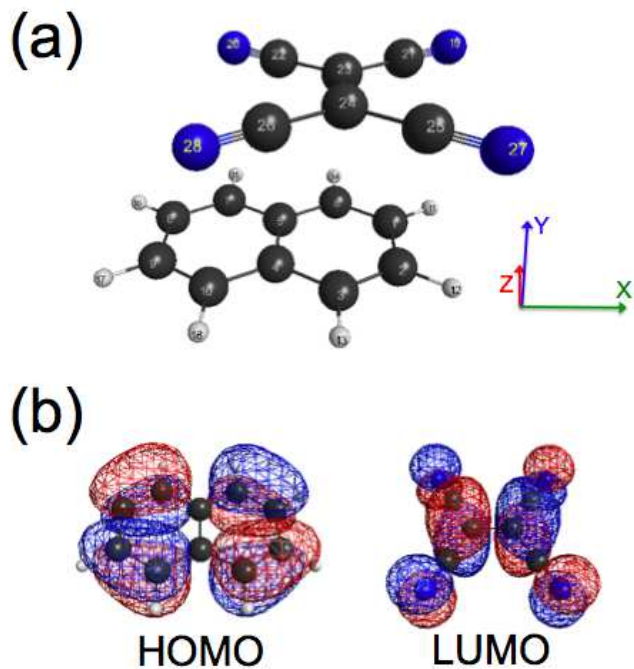}
\caption{
(a)
Schematics of the relative orientation of the monomers in a NPTL-TCNE dimer.
The Cartesian axes are also shown as an eye guide.
Atom number labels are also shown. 
The molecular planes of these two monomers are parallel to the X-Y plane.
\color{black}
The central C(23)-C(24) bond line of TCNE is parallel to the Y axis 
and this is placed on the C(3)-C(6) of the right aromatic ring in NPTL.  
\color{black}
Note that C(3)-C(6) line is also parallel to the Y-axis.
We modulate the distance of these two molecular planes 
as the difference in their Z-coordinates, namely, 
the distance of these two molecular planes.
\color{black}
(b)
HOMO of NPTL(donor) and LUMO of TCNE(acceptor). 
Orbitals were obtained both for two molecules as isolated systems. 
at the ab initio level of DFT/6-31Gd/CAMB3LYP.   
NPTL HOMO and TCNE LUMO energies are -0.2622 (Hartree) and -0.1367, respectively.
\color{black}
}  
\label{sys-nap-tcne}
\end{figure}

In the Fig. \ref{sys-nap-tcne}, 
we present the geometry configuration of this dimer 
to show the relative orientations of monomers. 
Each monomer was optimized at the DFT/6-31G(d) level of theory using 
the CAM-B3LYP exchange correlation functional.\cite{CAMB3LYP} 
The principal axis of NPTL is parallel to the X-axis, 
while that of TCNE is parallel to the Y-axis. 
The molecular planes of these flat molecules are parallel to the X-Y plane. 
Although we treat a dimer here, 
we partly considered the reported crystal data 
\cite{crystal-NTFL-TCNE-1967}
with respect to the relative orientation as explained in the figure caption.
Note that the relative orientation employed here gives rise to a non-vanishing overlap 
between the frontier orbitals, that is, the HOMO of NPTL and the LUMO of TCNE, 
which is included in the panel (b) in Fig. \ref{sys-nap-tcne}.

\color{black}
Throughout this article we use 
a primitive combination of a basis set 6-31G(d) and functional CAM-B3LYP 
because we focus on the validity of methodology. 
We note that that 
the dependency of the dynamics on the basis function and functional 
\color{black}
are important issue for charge transfer dynamics in 
a framework of DFT and TDDFT, which will be reported in our future articles.

%
%  result : charge migration dynamics between NPTL and TCNE under laser fields
%
\begin{figure}[th]
\includegraphics[width=0.85\textwidth]{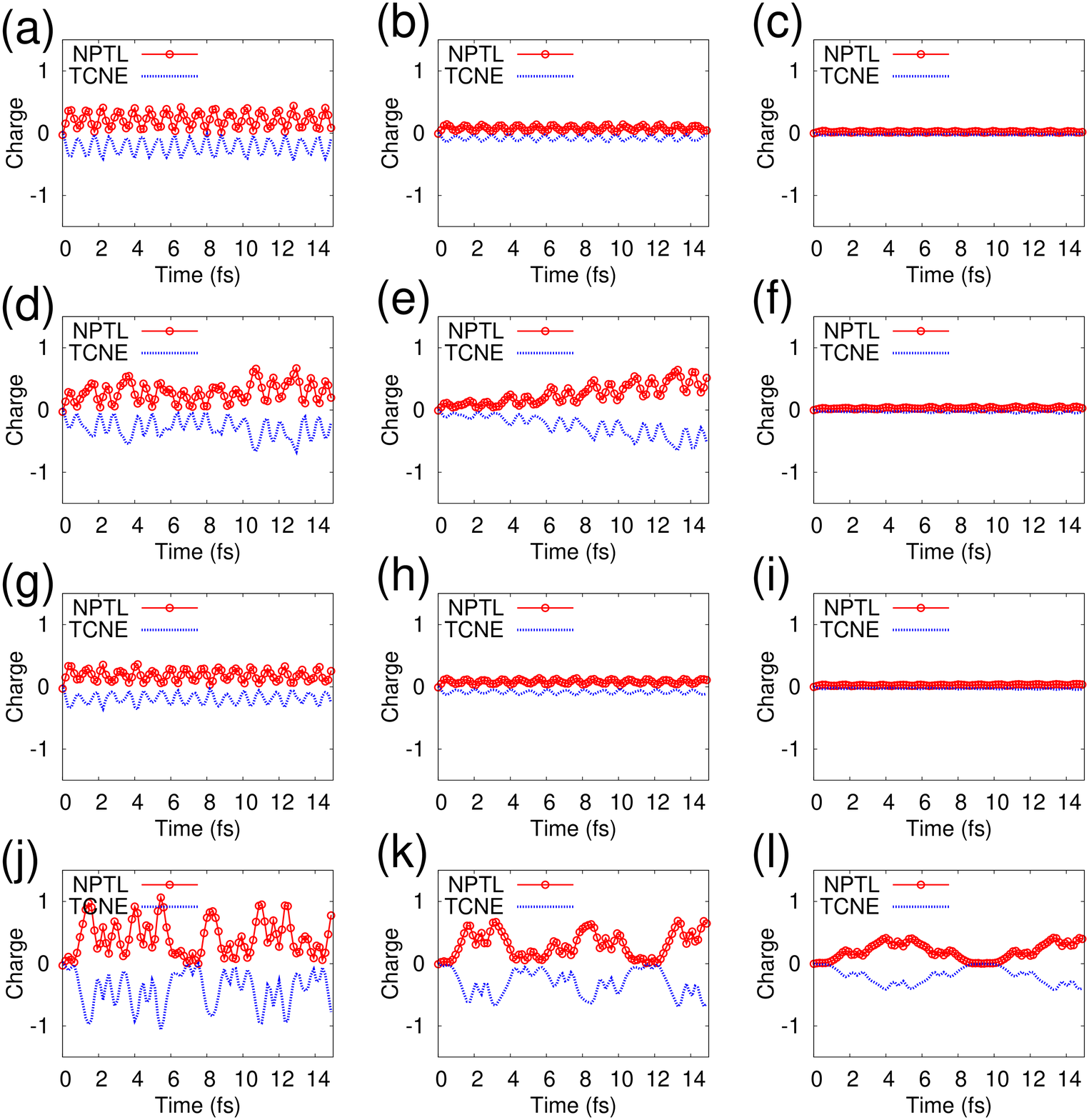}
\caption{
\color{black}
Charge migration dynamics in a NPTL-TCNE dimer.  
The panels include the cases with and without continuum light fields of 
a wave length of 700 nm and an electronic field strength of 0.02 a.u. 
The distances between the two monomer planes are 3.0, 3.5 and 4.0 \AA \,
in the left, middle and right columns, respectively. 
The three panels, (a/b/c), in the top row are the cases without any external light fields.
The other nine panels, (d--l), concern the cases with a light field. 
Normalized polarization vectors 
corresponding to the 2nd, 3rd and 4th rows are,
respectively, 
$(X,Y,Z)=(1,0,0)$, 
$(0,1,0)$,
and
$(0,0,1)$. 
\color{black}
}
\label{cs-nap-tcne-A}
\end{figure}

%
%  result : charge migration dynamics between NPTL and TCNE under laser fields
%
\begin{figure}[th]
\includegraphics[width=0.85\textwidth]{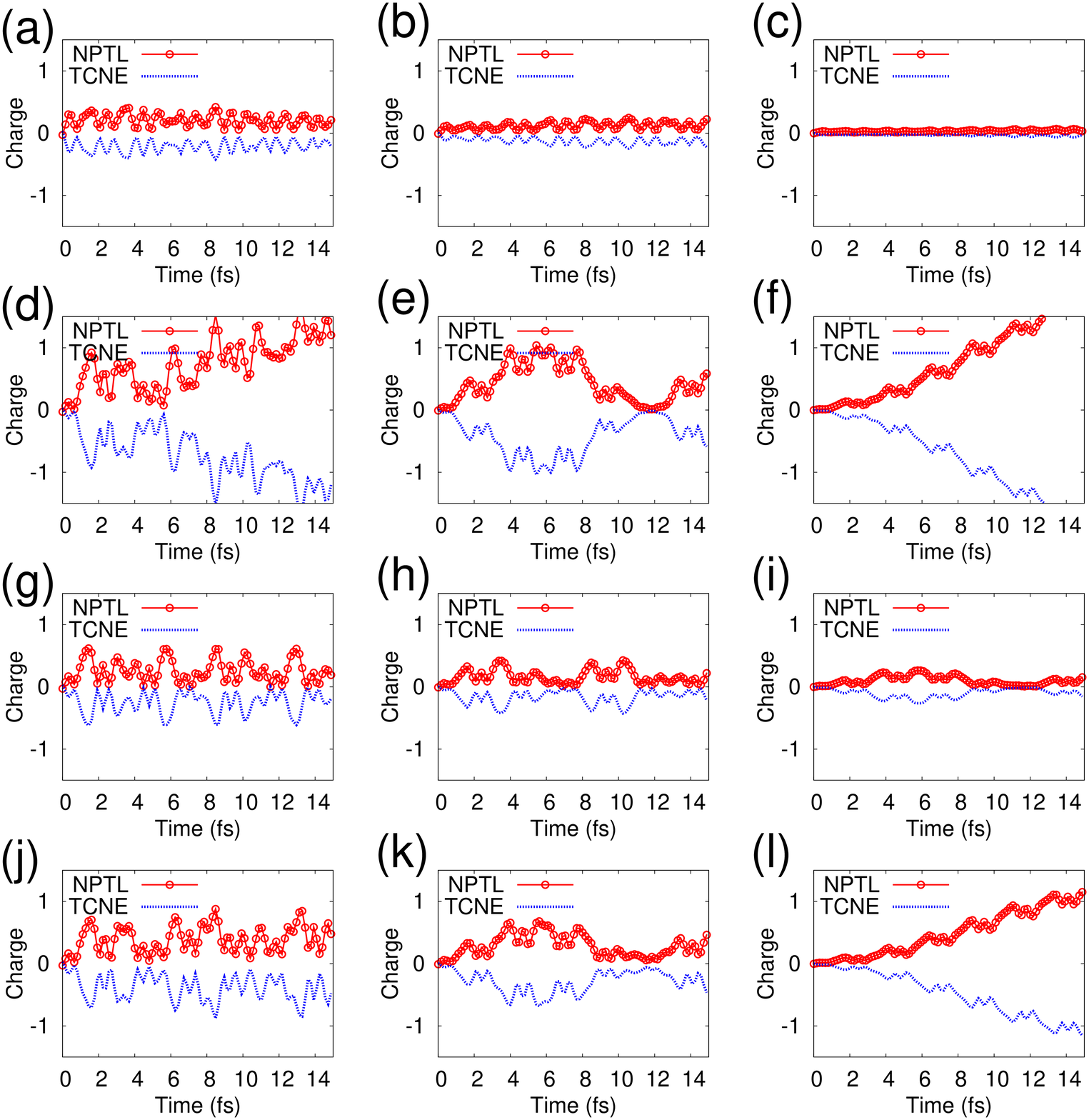}
\caption{
\color{black}
Charge migration dynamics in a NPTL-TCNE dimer.  
The panels include the cases with and without continuum light fields of 
a wave length of 700 nm and an electronic field strength of 0.02 a.u. 
The distances between the two monomer planes are 3.0, 3.5 and 4.0 \AA \,
in the left, middle and right columns, respectively. 
The twelve panels, (a--l), concern the cases with a light field. 
Normalized polarization vectors 
corresponding to the 1st, 2nd, 3rd and 4th rows are, respectively, 
$(1/\sqrt{2},1/\sqrt{2},0)$,
$(1/\sqrt{2},0,1/\sqrt{2})$,
$(0,1/\sqrt{2},1/\sqrt{2})$,
and 
$(1/\sqrt{3},1/\sqrt{3},1/\sqrt{3})$.
\color{black}
}
\label{cs-nap-tcne-B}
\end{figure}

By using this system, we examine 
(i)   the charge migration from the initial neutral electron filling for monomers 
      in the GD representation with and without a light field 
(ii)  the dependency of the monomer distance on the dynamics 
      with the different separations, (3.0, 3.5 and 4.0 \AA) \,  
and
(iii) the dependency of the field polarization directions on the results. 
%
% the way of arrangement of datain the figure
%
We summarize the results of the charge dynamics of the two monomers 
in Fig. \ref{cs-nap-tcne-A} and \ref{cs-nap-tcne-B}. 
\color{black}
The charges associated with the monomers were evaluated 
using Mulliken population analysis\cite{Szabo}
by using a density matrix of the whole system. 
The charges of monomers were simply evaluated 
as a sum of those of constituent atoms in them. 
\color{black}
The distance between the molecular planes increases from the left column to the right column, 
while the laser field conditions vary with respect to the rows shown in the figure. 

First we start from Fig. \ref{cs-nap-tcne-A}. 
% NO field
The top three panels, (a/b/c), 
present the results in the cases without a light field.  
Note that, initially, each monomer is neutral within the GDF representation.
From the time-dependent charges of the monomers we found that  
there was a small amount of charge transfer oscillation between these two monomers. 
Note that negative and positive charges in the figure 
correspond to the acceptance and release of electrons, respectively.
The amplitude is large when the distance is small, 
which is a reasonable result because the magnitude of molecular interaction 
depends qualitatively on the overlap of relevant orbitals.

% X
Next, we see the cases where the field has a polarization vector along the X-axis. 
These results are shown in the panels (d/e/f) in Fig. \ref{cs-nap-tcne-A}.   
\color{black}
In the panels (d) and (e), we find the sign of charge separation 
for the cases of distances of 3.0 and 3.5 \AA \,, respectively,
although not completely.
\color{black}
As shown in panel (d), for the shortest separation of 3.0 \AA, 
we found that the charge transfer was enhanced compared to the field-free case. 
The revival period after charge migration is about 8 fs. 
Interestingly, by increasing the separation to 3.5 \AA, 
the charge separation signature appears as shown in the panel (e), 
where the half period is 14 fs.
Note that the revivals are due to the fully coherent approach 
associated with a fixed molecular geometry. 
The observation here indicates that 
a reduction in the orbital overlap increases the time 
required to reach the peak of charge separation under an external field. 
In the other word, a very small separation is disadvantageous 
for meaningful charge separation between the donor and acceptor molecules, 
possibly arising from possible back-and-forward electron donation. 
However, as intuitively understood, a very large distance also 
reduces the efficiency of charge transfer because of the small interactions 
caused by the small overlaps of the relevant orbitals. 
As seen in the panel (f), 
\color{black}
which illustrates a case of a separation of 4.0 \AA, 
\color{black}
the large distance suppresses the orbital overlap and, thus, 
the charge transfer time becomes too large. 
In fact, the charge transfer is negligible at this time scale.

% Y
In contrast, 
the optically forbidden direction of light polarization for the system 
does not induce the charge transfer. In fact, 
as shown in the panels (g/h/i) in Fig. \ref{cs-nap-tcne-A}, 
which correspond to the polarization direction 
along the Y-axis, 
we found negligible charge separation,
similar to the field-free cases shown in (a/b/c) in Fig. \ref{cs-nap-tcne-A}. 

% Z 
Next, we see the results of the cases with a Z polarization of the light field 
which are shown in the panels (j/k/l) in Fig. \ref{cs-nap-tcne-A}.
Comparing these results with the previous cases, (a--i) in Fig. \ref{cs-nap-tcne-A},  
we found two clear different features in the charge migration dynamics.   
First, there is an increase in the number of oscillation 
in the envelope dynamics in (j) and (k), 
and, second, there is an enhancement in the moderate charge transfer 
at the largest distance, 4.0 \AA \, as shown 
\color{black}
in the panel (l).
\color{black}
As the distance increases from (j) to (l) 
the time-dependent charge behavior becomes smooth. 
%

%  more than two components 
Let us proceed to the cases with more than two components in the field polarization vectors, 
as shown in the panels (a--l) in Fig. \ref{cs-nap-tcne-B}.
We will see that the charge separations are significant 
in the cases including both X- and Z-components.
% X-Y
First, 
the results in cases including both X- and Y-components in the light polarization 
are presented in panels (a/b/c) in Fig. \ref{cs-nap-tcne-B}.
Comparing with the results shown in (d/e/f) in Fig. \ref{cs-nap-tcne-A}, 
we find that the addition of the Y-component reduces the charge separation. 
% X-Z
In contrast, the combination of X and Z components 
drastically increase the efficiency of charge separation,  
as shown in panels (d/e/f) in Fig. \ref{cs-nap-tcne-B}. 
Among the cases presented here, this combination gave rise to the best performance. 
\color{black}
It may be instructive to compare with the case including only X component. 
According to the discussion in the cases of (d/e/f) in Fig. \ref{cs-nap-tcne-A} 
with respect to the reduction of electron overlap with increase of the monomer distance,  
the difference in the degrees of charge separations between 
(k/l) in Fig. \ref{cs-nap-tcne-A} including only z component in the poralization vectors
and 
(e/f) in Fig. \ref{cs-nap-tcne-B} including x and z components  
indicate that the increase in the revival period and 
the emergence of the associated charge separation can be attributed 
to the suppression of electron back-donation  
caused by the addition of the X-component to the polarization vectors. 
\color{black}
% Y-Z
\color{black}
Interestingly, the Y-polarization of light field in this molecular configuration 
suppressed the charge separation which can be read from the cases with the combination 
of Y- and Z-components in the panels (g/h/i) in Fig. \ref{cs-nap-tcne-B}. 
\color{black}
% X-Y-Z
Then, how about a charge migration dynamics between monomers 
under the use of the polarization vector including all the components? 
We displayed the results for the case including three components, X, Y and Z 
in the polarization vectors of light 
in the three panels (j), (k) and (l) in Fig. \ref{cs-nap-tcne-B}  
with different distances of monomer planes, 3.0, 3.5 and 4.0 \AA, respectively. 
\color{black}
These panels show the signature of charge separation induced 
by the external light field.   
Differently from the first two panels (j) and (k) with smaller distances of planes, 
in the panel (l) with the largest distance of 4.0 \AA,   
we can see a clear charge separation in this time scale. 
\color{black}

%--- old : before referee comment 
%This can be considered as a kind of noise-induced effect
%associated with the numbers of components included
%in the light polarization vector.
%For example, Aspuru-Guzik {\it et al.} reported, through numerical investigations,
%that noise in the quantum subsystem can enhance 
%the efficiency of charge separation. \cite{env-assist-qw-eet}
%Xie {\it et al.} also found that
%increasing the number of bath modes coupled to the sub systems 
%associated with the charge migration 
%suppresses the oscillation between the donor and acceptor moieties 
%and assists the unidirectional flow of electrons.\cite{Lan}
%--- ref-comment: for 7/10/2017 comment
% This system is a kind of coupled multi level model. 
% 1) simple interpretation: different states are excited leading to different oscillation pattern
% 2) some oscillations are simply not covered in the restricted time range
% 3) there might be an electronic bath effect
%-- new: after comment
\color{black}
The understanding of the difference in charge separability 
with respect to the polarization vectors in the cases of largest distance, 4.0 \AA \,
corresponding to the panels (c/f/i/l) of Fig. \ref{cs-nap-tcne-B}   
are supported by the information of 
the dynamics of induced dipole moment vectors and their Fourier transformation 
associated with the optical responsibilities of the system.  
The propensity of optical transition and its magnitude for the monomers are key factors.  
This discussion is given in the supplementary material including detailed explanations 
accompanied with the data of the transition dipole moment vectors  
and localized orbital energies of monomers. 
Generally speaking, in the present demonstrations, 
the reduction and increase in the charge separation observed here 
by changing these multi components in polarization vectors
are attributed to a kind of electronic-bath effect 
associated with superpositions of electronic mode oscillations
having various time periods induced by simultaneous excitations of different states.  
\color{black}
%
%%                                                                                                 

%Although the ingredients associated with the electron dynamics here 
%are different from those in previous reports by Aspuru-Gzuik and Xie,
%the tendency observed in the present charge separation dynamics 
%under an external field is qualitatively consistent with those statements 
%on replacing the roles of vibrational motion with that of external radiation field.

%
%
%
Through the investigation of the simple donor-acceptor system, 
we obtained the following two findings; 
(I)
a time-dependent approach coupled with 
the group diabatic representation yields 
a microscopic information on electron properties such as charge dynamics,  
which can not be obtained by 
a static analysis or nuclear wave packet calculation
\color{black}
( The statement above does never deny static and nuclear wave packet approaches. 
  In fact, they can offer a useful and high level information 
  on electronic properties of charge transfer and nuclear quantum effect.  )
\color{black}
, and    
(II)
the initiation dynamics of the charge separation are 
sensitively dependent on the distances of the constituent monomers 
and the optical properties associated with the donor-acceptor system, 
which also cannot be extracted from a static analysis. 

Of course, the effects of electronic nonadiabaticity, as well as 
molecular dynamics, are also important for the charge separation dynamics.  
However, here, we have not touched these issues for clarity. 
They will be discussed and reported in our future articles.

\subsection{Unpaired electron dynamics: 20-mer ethylene}

In the next demonstration, 
\color{black}
we focus on the unpaired electrons in excited electron dynamics.  
\color{black}
The diffusion dynamics of the excitons
over weakly interacting monomers are examined.  
We examine the differences between results 
obtained with and without the group localization procedure.  
The group localization scheme provides us 
with a clear view of the migration dynamics 
of local excitons in the monomers 
in the presence and absence of a laser field.

\subsubsection{Effective unpaired electron (EUPE)}

First, we define an unpaired electron as used in this paper.
An effective unpaired electron density matrix was constructed 
by extracting the components corresponding 
to an occupation of almost one (half filling) 
of the natural orbitals from the original density: 
\cite{unpaired,nagashima2012}
\begin{align}
\hat{\rho}^\mathrm{EUPE} 
= \sum_i | \phi^\mathrm{NO}_i \rangle n_i(2-n_i) \langle \phi^\mathrm{NO}_i |, 
\label{eupe-dmat}
\end{align}
where 
$ \hat{\rho}| \phi^\mathrm{NO}_i \rangle = n_i | \phi^\mathrm{NO}_i \rangle $ 
with $\phi_i^\mathrm{NO}$ and $n_i$ being the {\it i}-th natural orbital and 
its corresponding natural population, respectively. 
\color{black}
Here, $\hat{\rho}$ 
\color{black}
denotes 
\color{black}
a one-electron density operator for 
the whole system.
\color{black}
This treatment enables us to obtain an information concerning the  
polyradical features of the complex system. \cite{b12-nad}

\color{black}
For example, in the GD representation case, 
quantities of unpaired electrons for monomers are evaluated as follows: 
\begin{itemize}
\item[1] Diagonalize electron density matrix, $\underline{\underline{\rho}}^{\textrm{GD}}$,  
   in the group diabatic representation, 
 $
 \underline{\underline{\rho}}^{\textrm{GD}} 
 \equiv 
 \underline{\underline{U}}  
 \,
 \underline{\underline{n}}
 \,
 \underline{\underline{U}}^{\dagger}
 $
 where $\underline{\underline{U}}$ and  $\underline{\underline{n}}$
 are a unitary matrix and diagonal natural population matrix. 
\item[2] Create a unpaired electron density matrix, 
 $
 \underline{\underline{\rho}}^{\textrm{GD:EUPE}}
 \equiv 
 \underline{\underline{U}}
 \,
 \underline{\underline{n}}
 \left( 2 \underline{\underline{1}} - \underline{\underline{n}} \right)
 \,
 \underline{\underline{U}}^{\dagger}
 $
\item[3]
 Convert $\underline{\underline{\rho}}^{\textrm{GD:EUPE}}$ into the form in AO representation  
 and perform the Mulliken population analysis for monomers. 
\end{itemize}
\color{black}

\subsubsection{System information on the 20-mer ethylene}

\begin{figure}[th]
\includegraphics[width=0.85\textwidth]{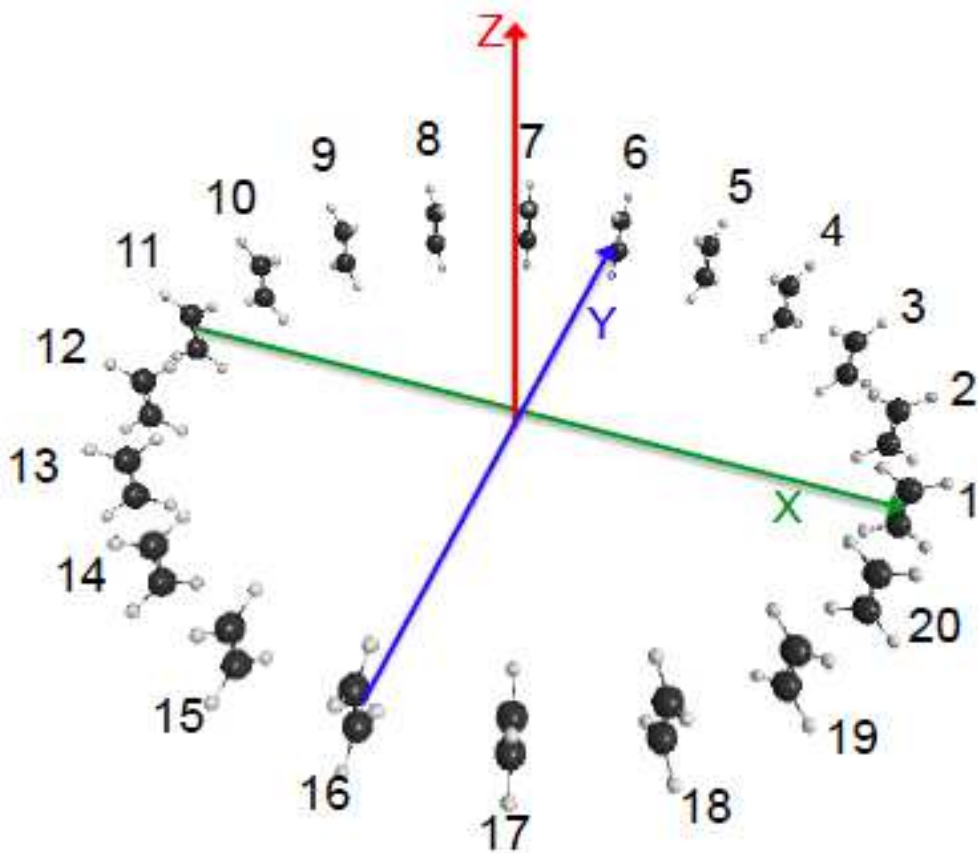}
\caption{Schematics of the relative positioning of a 20-mer circle of ethylene molecules.
The ethylene monomers are identified here by the attached numbers.  
The Cartesian axes are also shown to guide the eyes of readers.}  
\label{sys-twenty}
\end{figure}

For the assessment of the present method, 
we treated an aggregated system consisting of a circle of twenty ethylene molecules,
as shown in Fig. \ref{sys-twenty}. 
The centers of mass of the monomers are uniformly placed on 
the circle, which has a radius being 12 \AA. 
The C-C lines of planer ethylene molecules are aligned 
vertically to the circle plane. 
The normal vector of each ethylene is parallel 
to the tangent vector of this circle at the position of the monomer.

The geometrical structure of ethylene was determined 
at the CAM-B3LYP/6-31G(d) level of theory. 
We also employed the same level of ab-initio calculation 
for the 20-mer ethylene system as that used for ethylene monomer.

\subsubsection{Exciton migration dynamics}

%
% Es = 0.005(a.u.)  <-->  I = 3.75 x 10^11 (W/cm^2)
% Es = 0.01(a.u.)   <-->  I = 3.50 x 10^12 (W/cm^2)
% Es = 0.02(a.u.)   <-->  I = 1.40 x 10^13 (W/cm^2)
%

\begin{figure}[th]
\includegraphics[width=0.75\textwidth]{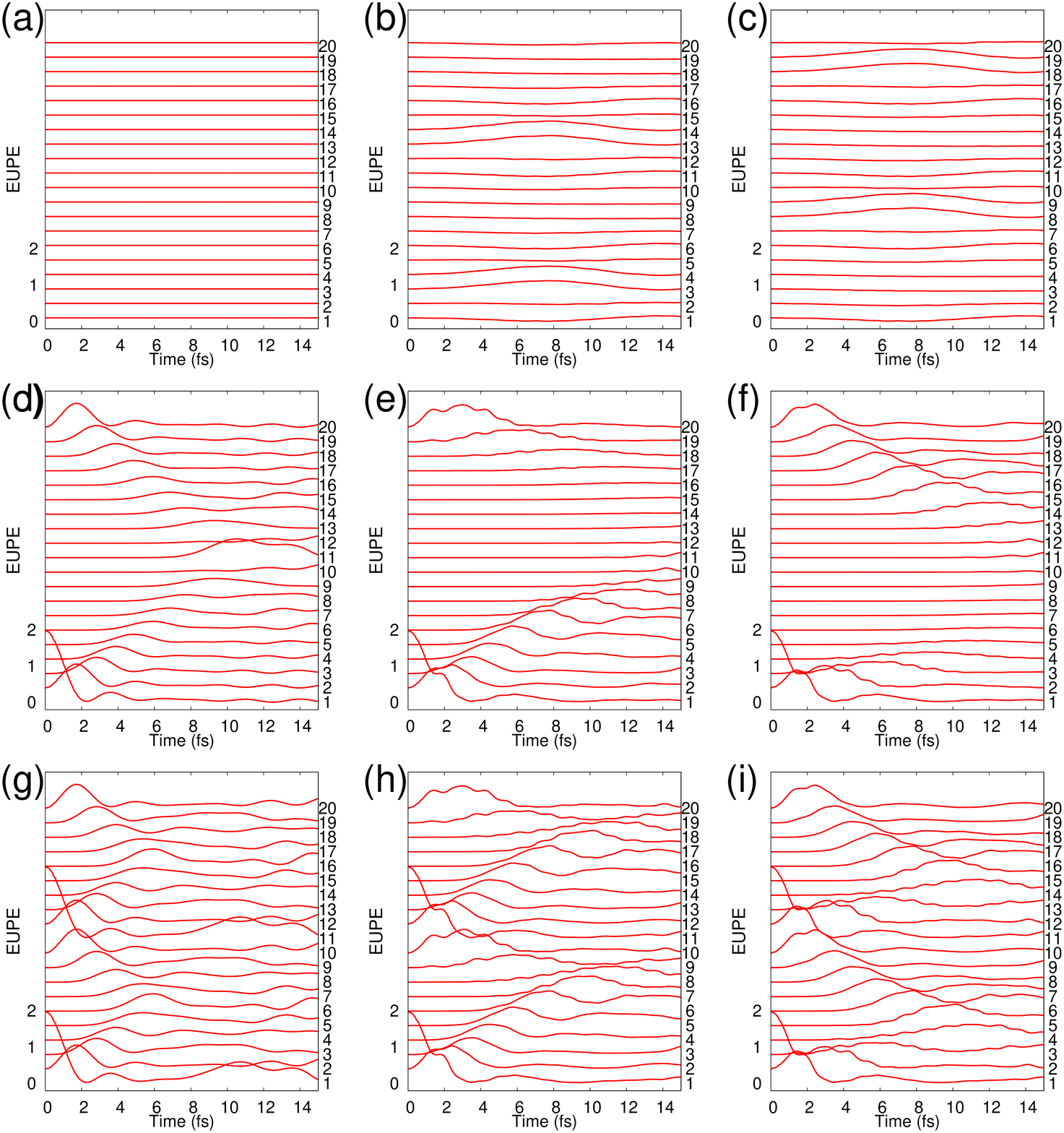}
\caption{
\color{black}
Time-dependent behaviors of quantities of unpaired electrons 
\color{black}
in the 20-mer circle of ethylene molecules.
\color{black}
Unpaired density matrix for the whole system, Eq. (\ref{eupe-dmat}),  
was applied to Mulliken population analysis for group monomers throughout (a-i).  
\color{black}
In the upper three panels, (a/b/c), 
no group diabatization was employed; that is, 
the whole system is treated as one monomer and   
the initial state corresponds to a single electronic excitation 
from the local HOMO to the LUMO orbital. 
In contrast, panels (d/e/f) show results of the  group diabatic localizations 
and  only 1-st monomer is initially excited from local HOMO to LUMO orbital   
while 1st and 11th monomers are initially excited in the panels (g/h/i)
with the group diabatization procedure. 
The three panels (a/d/g) correspond to the cases without the light field 
while the other panels concern light-electron couplings. 
The parameters of the light fields are as follows;  
the wavelength and strength are 700 nm and 0.02 atomic unit (a.u.), respectively. 
and normalized polarization vectors are 
$(X,Y,Z)=(1/\sqrt{3},1/\sqrt{3},1/\sqrt{3})$ 
for (b/e/h) and $(1/\sqrt{3},-1/\sqrt{3},1/\sqrt{3})$ for (c/f/i).  
The horizontal and left vertical axes are the time in 
\color{black}
femtoseconds  
\color{black}
and 
\color{black}
the quantity of unpaired electrons for monomers, 
\color{black}
respectively. 
The numbers shown on the right vertical axis are the monomer labels. 
See the main text for details of these plots.}
\label{eupe-nodiv-div-wf-wof-sym-pol}
\end{figure}

Here, we show that the group diabatic representation is an advantageous
for investigating the exciton migration dynamics in molecular aggregate systems 
containing weak interactions between monomers. 

First, we present the results in the cases without any group diabatization.  
Note that, here, we employ $N_g=1$ in the GDF representation. 
This means that the 20-mer ethylene system was treated as one monomer system.
Initially, the entire system was excited from the HOMO to the LUMO and 
the corresponding electron density matrix was prepared.
\color{black}
We note that the treatment of the total system as a monomer is rather natural.
This first presentation treating a whole system as a monomer is aimed 
to compare with the local excitation analysis by using group diabatic scheme 
performed later in this article.  
This means to see how the introduced method is useful in 
the local diabatic analysis in a time domain, which is the main topic in this article.
\color{black}
The panels (a-c) of Fig. \ref{eupe-nodiv-div-wf-wof-sym-pol}  
summarizes the results for the cases without any group diabatization.

For this initial electronic density matrix, 
two types of dynamics simulations were carried out:  
One with and one without an external optical field.

\color{black}
In the cases with light fields, as model cases of modification 
of excited electron transfer stirred by solar light, 
\color{black}
we employed the same parameters for the continuum light field 
except for the polarization vectors, which were used as 
those in the case of the NPTL-TCNE dimer.  
The details of parameters are included in the figure captions. 
The polarization direction vectors of the light field used here 
have in-plane and out-of-plane components with respect to the plane on which 
the center of masses of the ethylene monomers are positioned. 

Note also 
that the symmetry of the system is broken in a case accompanied with this light field, 
for example, symmetry operations with respect to the X-Z plane 
and this will also be true for the induced exciton migration.  

The number of unpaired electrons in each monomer was evaluated 
via Mulliken population analysis for the unpaired electron density, 
Eq. (\ref{eupe-dmat}).
For clarity, we have used the label number of monomer 
as the origin in plotting the time-dependent data sequences. 

As shown in the panel (a), in the absence of a light field, 
the unpaired electrons are distributed uniformly over 
the twenty monomers and no migration occurs. 
This indicates that the unpaired electron affinities are balanced among these moieties 
in this setting of calculation. 
In contrast, unpaired electron dynamics between the monomers appear weakly 
in the presence of an external radiation field, 
as shown in the panel (b) and (c).
The field polarization vectors in these two cases, 
are related by inversion with respect to the X-Z plane.    
Thus, the patterns of unpaired electron dynamics is 
in an inverse relationship between these two cases for the X-Z plane. 

From the resultant uniform excitation of the monomers shown in (a--c), 
the initial excitation of the entire system is not so useful 
for the analysis of the time dependent monomer interactions. 
The difficulty in the interaction analysis without group diabatization is resolved 
by applying a group diabatized representation and 
the associated local initial excitation explained in the Sect. \ref{sect-theory}. 
Now, let us proceed to trial calculations 
using the group diabatic representation discussed in this article.

%
%  eupe migration with using GD rep: hlex 1st only
%

Next, we examine the case where initially 
only one ethylene molecule is excited from the HOMO to the LUMO 
in the localized canonical orbitals of this moiety.
Here, group localizations are carried out by using all the monomers.
This means that we employed $N_g=20$ in the GDF representations.  
The results are presented 
in the panels (d)--(f) of Fig. \ref{eupe-nodiv-div-wf-wof-sym-pol}.

As seen in the panel (d), 
in the absence of an external radiation field, 
the highly localized unpaired electrons in the 1-st monomer are 
separately transferred to 11-th monomer at 10 fs, accompanied by a moderate 
broadening in the distribution over the monomers.
Symmetry with respect to the X-Z plane was also observed. 

In the case with the external light field of the panel (e), 
we found that the exciton transfer becomes slow 
associated with the field induced change in the interactions between monomers. 
The symmetry of the migration dynamics of the unpaired electrons 
with respect to the X-Z plane is clearly broken 
by the light polarization direction.  
We also present the results, in the panel (f), 
of the calculation with a symmetric light polarization 
compared to the case of (e) with respect to the X-Z-reflection plane. 
In this case of (f), the symmetry breaking is the inverse of that (e).

%
%  eupe migration with using GD rep: hlex 1st and 11-th 
%
As a final example of unpaired electron dynamics, we examine the cases of 
initial two-site single excitations from the local HOMO to the LUMO  
for the 1-st and 11-th monomers, for which results are included 
in the panels, (g), (h) and (i). 
The conditions of the light fields used in the panels (g/h/i) are 
the same as those in (d/e/f) in this ordering, 
for which details are included in the figure caption. 
As shown in panel (g), 
there is a clear bifurcation and confluence of two localized excitions 
accompanied with moderate dispersion. 
These dynamics can be modified by an external light field. 
The panels, (h) and (i), show the two cases with polarized directions 
related by reflection with respect to the X-Z plane. 
From these data, we can observe that 
the directions of rectification of the exciton flows induced by the light field 
are inverse with respect to the X-Z plane for these two cases. 
\color{black}
The difference in the Fourier spectrum and time-dependent induced dipole moment vectors 
between one and two initial local excitation cases are summarized in the supplementary material, 
where the readers can find the information on the mode of electronic motions  
dominantly triggered by the initial local excitation and following optical response 
of electrons.  
\color{black}
\color{black}
As seen in the demonstration of initial two-Frenkel-exciton case, 
the method can safely describe the details of the exciton migration dynamics over the monomers. 
This is also the case for the one exciton case as shown above.
We give schematic movies in the supplemental materials 
for the two excitation cases with and without the light field, 
where readers can see the unpaired electron wave packet dynamics, 
especially the extent of the delocalization of packets, time scale in it 
and rectification feature associated with external light field.  
\color{black}

%
%  charge migration with using GD rep: hlex 1st only vs 1st-11th 
%

\begin{figure}[th]
\includegraphics[width=0.8\textwidth]{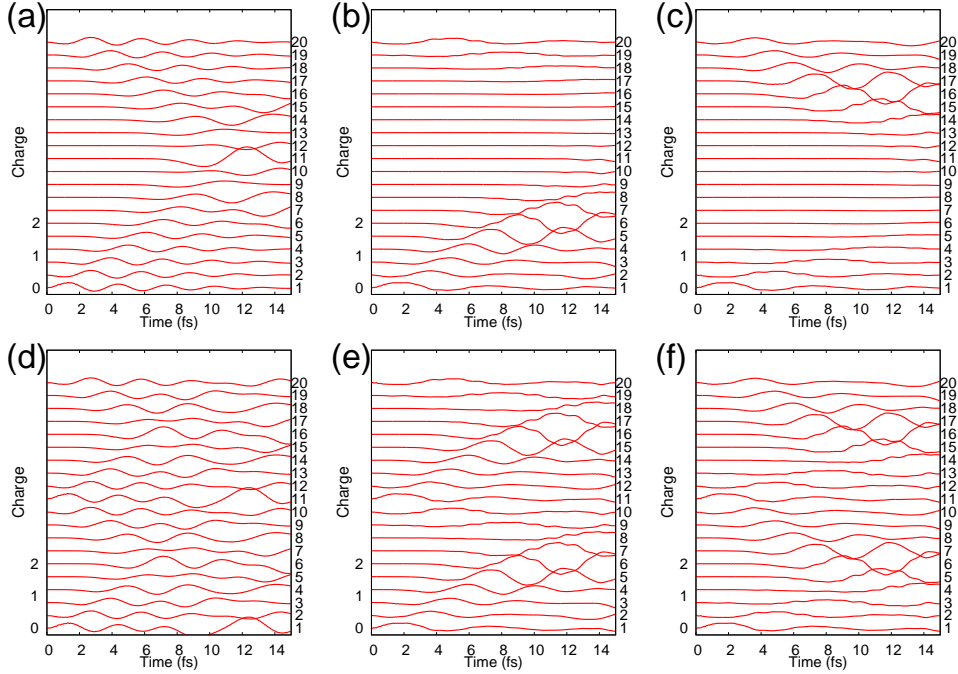}
\caption{
\color{black}
Time-dependent Mulliken charge in the 20-mer circle of ethylene molecules.
Panels (a/b/c) show results of the group diabatic localizations 
and only 1-st monomer is initially excited from local HOMO to LUMO orbital   
while 1st and 11th monomers are initially excited in the panels (d/e/f)
with the group diabatization procedure. 
The three panels (a/d) correspond to the cases without the light field 
while the other panels concern light-electron couplings. 
The parameters of the light fields are as follows;  
the wavelength and, strength are 700 nm, 0.02 atomic unit (a.u.), respectively. 
and normalized polarization vectors are 
$(X,Y,Z)=(1/\sqrt{3},1/\sqrt{3},1/\sqrt{3})$ 
for (b/e) and $(1/\sqrt{3},-1/\sqrt{3},1/\sqrt{3})$ for (c/f).  
The horizontal and left vertical axes are the time in femtoseconds  
and the scaled unpaired electron quantity for the monomers, respectively. 
The numbers shown on the right vertical axis are the monomer labels. 
\color{black}}
\label{charg-wf-wof-sym-pol}
\end{figure}

\color{black}
So far we examined unpaired electron dynamics. 
How about the dynamics of charge of local sites ?
In Fig. \ref{charg-wf-wof-sym-pol}, 
we showed the result of Mulliken charge migration dynamics over twenty monomers. 
The panels of (a)-(c) display the result of the dynamics starting from initial density matrix 
corresponding to an excitation of 1st monomer from local HOMO to LUMO 
while in the panels of (d)-(f) 1st and 11th monomers are initially excited in the same way. 
Here the diabatic group number, $N_g$, is twenty. 
The field parameters of (a)-(f) in Fig. \ref{charg-wf-wof-sym-pol} 
are the same as those of (d)-(i) in Fig. \ref{eupe-nodiv-div-wf-wof-sym-pol} in this order.
The plotting manner are the same as that of Fig. \ref{eupe-nodiv-div-wf-wof-sym-pol}.

By comparing the charge migration dynamics in 
the panels (a)-(f) in Fig. \ref{charg-wf-wof-sym-pol}, 
with the time dependent unpaired electrons of local sites shown in 
the corresponding panels (d)-(i) in Fig. \ref{eupe-nodiv-div-wf-wof-sym-pol}, 
we find the following features and obtain the results from them: 
\begin{itemize}
\item [1]
As a summary, the charge migration clearly seen 
in Fig. \ref{charg-wf-wof-sym-pol} indicates that 
the initially prepared local excitons are characterized by a charge transfer type of exciton. 
The charge density wave packets slightly delocalized over two or three monomers 
propagate spatially along the molecular chains in the molecular aggregate system, 
and drive the charge migration over twenty ethylene monomers.
We can consider the present electron migration dynamics 
as a charge transfer excition type of dynamics. 
\item [2] 
In fact, as seen in the panel (a) of Fig. \ref{charg-wf-wof-sym-pol} 
with the one site excitation and without the light field, 
the initial Mulliken charge of the initially excited 1st ethylene site is zero 
and this monomer, soon after the onset of the dynamics,  
obtains the positive charge from the nearest two 2nd and 20-th sites. 
Then this positive charge stored in 1st site in turn propagates  
in a splitting manner via 2nd and 20th monomers  
toward the 11th monomer at the counter position with respect to the 1st monomer. 
This two charge density wave packets encounter at 11th around the time of 12 fs 
accompanied with a constructive interference. 
\item [3]
The panel (d) of  Fig. \ref{charg-wf-wof-sym-pol} 
for the case of initially prepared two excitations 
show the same feature as (a) of the same figure 
except for the double numbers of spawned charge density wave packets. 
\item [4]
Compared to the panels of (e), (f), (h) and (i),
in Fig. \ref{eupe-nodiv-div-wf-wof-sym-pol} with respect 
to the light field case with initial excitations, 
the corresponding panels (b), (c), (e) and (f) 
of in Fig. \ref{charg-wf-wof-sym-pol} show the same shapes of tracks 
and rectification tread for charge density wave packets. 
\item [5]
The packets have the width over two or three sites during dynamics, 
which characterize the exciton delocalization. 
This means that the initially prepared Frenkel like exciton at each local site 
was converted into stable delocalized excitons accompanied with a charge 
during the dynamics.   
\end{itemize}

%Additionally, the results obtained in the cases of light field  
%mean that the external perturbations can regulate the excitation energy transfer. 
%This can be considered as a control over the excited electron dynamics 
%by using an external light field. 
Though these calculations include rather artificial initial setting of local excitation, 
we can extract inherent dynamical trend in this molecular aggregated system 
between local excitation and local charge by using GDF electron dynamics method. 
\color{black}

Thus, 
through the dynamics calculations with and without 
the group diabatization representation, various local excitations and laser fields,  
we found that this analysis has a utility to obtain an information on 
a dynamical propensity of excited electrons 
in aggregated systems having sparse networks of the interactions 
between the constituent monomers.

\section{Summary}
\label{sect-summary}

In this paper, 
we have introduced a calculation scheme for excited electron dynamics
based on the group diabatic Fock representation.
We verified that this GDF electron dynamics method  
allows for the concise description and analysis of the  
excited electron migration dynamics in molecular aggregate systems.   
This was assessed by using 
an elemental light energy conversion material 
made of electron donor and acceptor molecules 
and a one-dimensional system consisting of circularly oriented monomers.  
The dynamics were characterized 
according to the inherent gradient of electron affinities 
among the local molecular groups under the employed conditions 
of initial excitations and external laser-electron couplings. 

The present scheme is advantageous for the future ab initio modeling 
of excited electron migration dynamics in a moderately large system. 
This is because the GDF representation provides 
a clear strategy for the extraction of the active orbitals 
at each local group site 
by setting the orbital energy range related to 
the excited electron transfer under consideration. 

In this aspect, the promising scheme recently developed by 
Shimazaki, Kitaura, Fedrov and Nakajima  as a fragment type 
\color{black}
dual-layer 
\color{black}
self consistent field theory for the treatment of a large sparse system 
\cite{GMO} will play an important role in the exploration of 
the roles of the structured but complex interactions 
of many types of molecular aggregate systems.
In fact, the present work is partly inspired by the work mentioned above 
and can be combined with it by replacing the localization scheme with 
\color{black}
\textit{e.g.} Boys\cite{Boys}, Pipek-Mezey\cite{PM} methods and so on. 
\color{black}

The external, as well as internal, fields imposed on the systems 
affect the excited electron dynamics and energy transfer.  
Therefore, the molecular motion associated with nonadiabatic transition 
among complex excited states are also important 
at longer timescales than that considered in the present article. 
\cite{env-assist-qw-eet} 
Furthermore, in systems containing metal atoms, 
the spin-orbit couplings become important for excited state dynamics 
involved with inter-system crossings.
\cite{Daniel-ACR-SOC-QM-DYN,Fedrov-AIMS-SOC}
We will report these issues in future articles by 
using the extended version of the present GDF electron dynamics scheme.

\section{Supplementary Material}
\color{black}
See supplementary material for Fourier analysis of induced dipole moment 
of the NPTL-TCNE dimer and 20-mer ethylene systems,
the fundamental properties of 
transition dipole moments as well as localized orbital energies for monomers  
and the schematic movies of dynamics of local unpaired electrons in the latter system. 
\color{black}

\begin{acknowledgements} 

The authors thank Dr. Michio Katouda and Dr. Keisuke Sawada 
for their advises, discussions and technical support 
concerning the use of the NTChem code and parallel calculations. 
We are grateful to Prof. Kazuo Kitaura and Prof. Unpei Nagashima 
for the valuable information on the molecular interaction concept 
in terms of molecular orbitals.  
We also appreciate the discussions with Dr. Tomomi Shimazaki, 
Prof. Yuzuru Imamura, Prof. Motomichi Tashiro and Prof. Mikiya Fujii 
regarding the charge separation mechanism in solar cell systems.
This research was supported by MEXT, Japan, 
``Next-Generation Supercomputer project'' 
(the K computer project) 
and ``Priority Issue on Post-K computer'' 
(Development of new fundamental technologies 
for high-efficiency energy creation, conversion/storage and use).
The part of computations in the present study was performed 
using Research Center for Computational Science, Okazaki, Japan.

\end{acknowledgements}

% ===============

%
%
% supplementary materials
%
%

\clearpage

{\Large
Supplementary material for 
``A quantum dynamics method for excited electrons in molecular aggregate system 
using a group diabatic Fock matrix'' 
}
\normalsize

\clearpage

%%%%%%%%%%%%%%%%%%%%%

%\begin{figure}[th]
%\includegraphics[width=1.0\textwidth]{Fig-GDFrevise1-S1.eps}
%\caption{Absorption spectrums of NPTL-TCNE.}
%\label{spectrum-NPTL-TCNE}
%\end{figure}

\section{Suppl.1: Mode analysis of electron dynamics in NPTL-TCNE dimer}

\begin{figure}[th]
\includegraphics[width=0.7\textwidth]{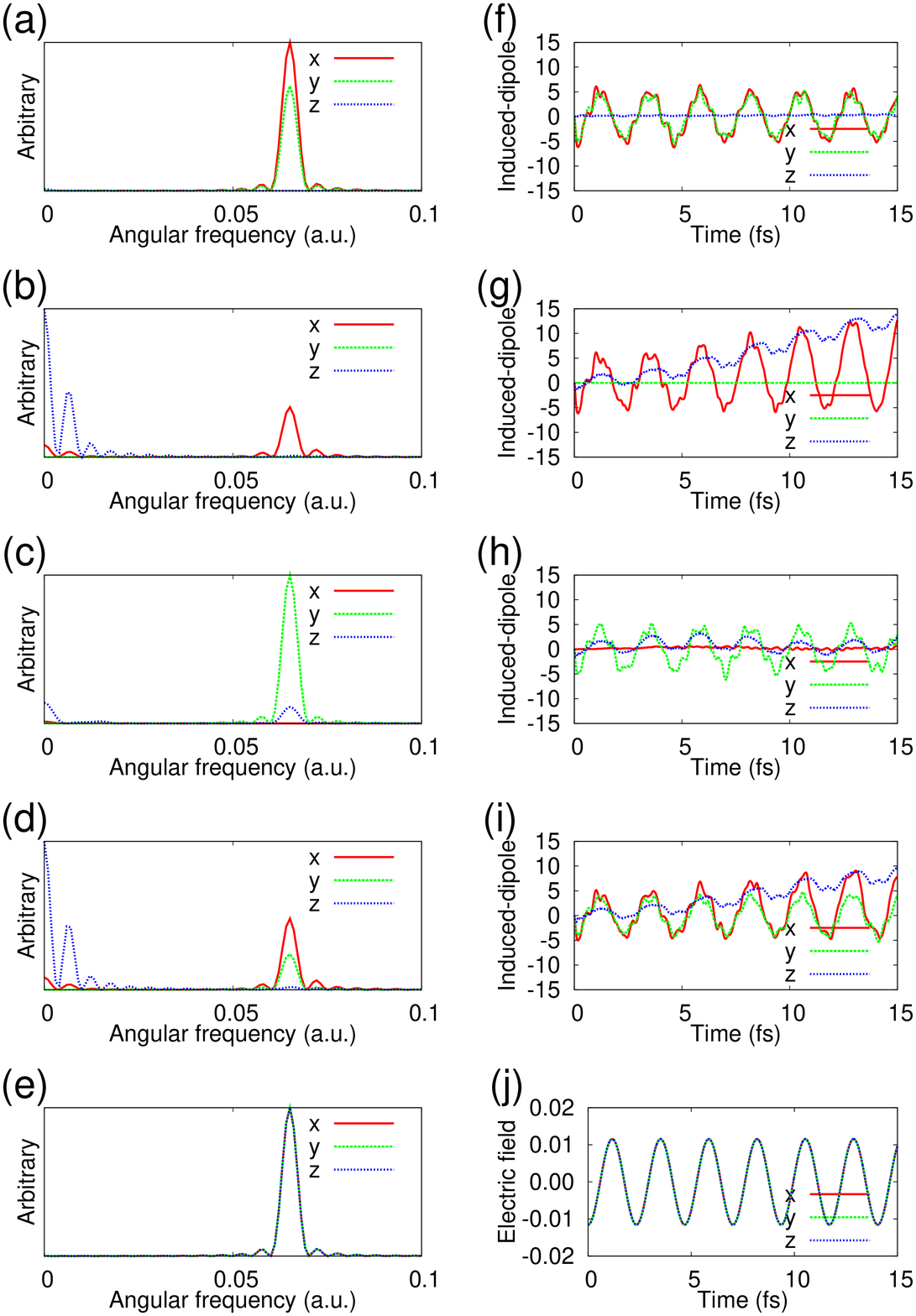}
\caption{Electronic mode analysis in the cases of a dimer NPTL-TNCE system 
with group diabatic ground state, which are included in the top four raws.  
The distance of two molecular plane is 4.0 \AA. The system is shined by the continuum light 
wave with 700nm wavelength (corresponding to one photon energy, 0.065 a.u.) 
and normalized polarization vector of (x,y,z) = 
$(1/\sqrt{2},1/\sqrt{2},0)$, $(1/\sqrt{2},0,1/\sqrt{2})$, $(0,1/\sqrt{2},1/\sqrt{2})$ 
and $(1/\sqrt{3},1/\sqrt{3},1/\sqrt{3})$, 
for the panels of (a/f), (b/g), (c/h) and (d/i), respectively. 
The panels (a/b/c/d) are the Fourier transformations of the time sequence of 
induced dipole moment of the whole system which are respectively 
shown in the panels (f/g/h/i).
(e) is the Fourier transformation of the continuum light field 
used in the panels of (d) and (i) 
while (j) is the electric field in a time domain corresponding to (e).}
\label{FT-RT-ind-dip-Evec-dimer-NPTL-TCNE}
\end{figure}

In  Fig. \ref{FT-RT-ind-dip-Evec-dimer-NPTL-TCNE}, 
we summarized the mode analysis of GDF electron dynamics in the dimer NPTL-TCNE system 
in the main text of the article for the cases including the light field 
having 700nm wave length corresponding to one photon energy 0.065 Hartree, 
and normalized polarization vectors of (x,y,z) = 
$(1/\sqrt{2},1/\sqrt{2},0)$, $(1/\sqrt{2},0,1/\sqrt{2})$, $(0,1/\sqrt{2},1/\sqrt{2})$ 
and $(1/\sqrt{3},1/\sqrt{3},1/\sqrt{3})$.  
Note that this light field is rather non-resonant 
to the typical local electronic transitions because the energy gaps of local HOMO-LUMO 
being  0.0010-(-0.2721)=0.2731 for NPTL  and -0.1278-(-0.3815)=0.2537 for TCNE
are far larger than 0.065. 
The results in the cases with the distance of molecular planes being 4.0 \AA \,
are selectively shown, because they include key information of 
the response dynamics of electrons with respect to the light polarization vectors 
having the elements more than two in this molecular orientation against for 
the laboratory frame in the present demonstration.  

Here, the time sequences of induced dipole moment and the Fourier transformation of them 
are displayed. For the light field, only the case of the polarization vector, 
$(1/\sqrt{3},1/\sqrt{3},1/\sqrt{3})$, is shown for the guide of eye 
to show the frequency and time dependent behavior of the external radiation fields. 
An induced dipole of total system used for analysis was defined 
as the difference of the time dependent dipole moment 
measured from the dipole moment at the initial simulation time, 
$\mu_k^{\textrm{ind}}(t) = \mu_k(t) - \mu_k(0) $ with $k$ running over x, y and z. 
The broadenings of the peaks in FT spectrums are caused 
by the finite resolution associated with the simulation time, 15 fs.  

The height for each component in FT spectrum at the peak of 0.065 a.u. 
in the frequency domain shown in the left column
has an information on the extent of the response of electrons 
for the polarization component in light field. 
The panels tell us how the electrons respond to the light field, 
with respect to the induced dipole moment. 

We can see that the light fields 
having the polarization vector including z component 
cause the gradual increase of z component in dipole moment as 
shown in the panels of (g) and (i) except for the panel (h) of the y-z pair polarization. 
The low frequency peaks far below 0.065 in the panels (b) and (d)
correspond to the slow increase of z component in the dipole moment associated with 
the unidirectional electron migration between monomers along z axis 
as shown in the panels of (g) and (i), respectively. 
We can see again that z component in the light field is essential for the electron transfer. 
In fact, there is no growth of dipole moment in the z direction 
in the panels (a) and (f) for the case including x and y components 
in the polarization vector but not having z component in it.  

Interestingly, y component in light polarization vector suppresses 
the electron transfer along z axis 
caused by z component of the light polarization vector.  
As found in the panels (c) and (h) including y and z components in the polarization vector, 
the response of electrons along y axis is clearly superior to that along z axis. 
This indicates that for this field the system has 
a strong electronic transition property along y axis,   
which resultantly prevents the opportunities of the electron transfer along z axis 
parallel to the normal vectors of two molecular planes.  
This picture is supported by the fact that 
for the NPTL monomer {\it as the electron donner}
the local HOMO-LUMO transition dipole moment vector has 
large values for the y component and zero values for x and z component 
as seen in Tab. \ref{tab-NPTL}.

\section{Suppl.2:Mode analysis of electron dynamics in 20-mer circle of ethylene molecule}

\begin{figure}[th]
\includegraphics[width=0.70\textwidth]{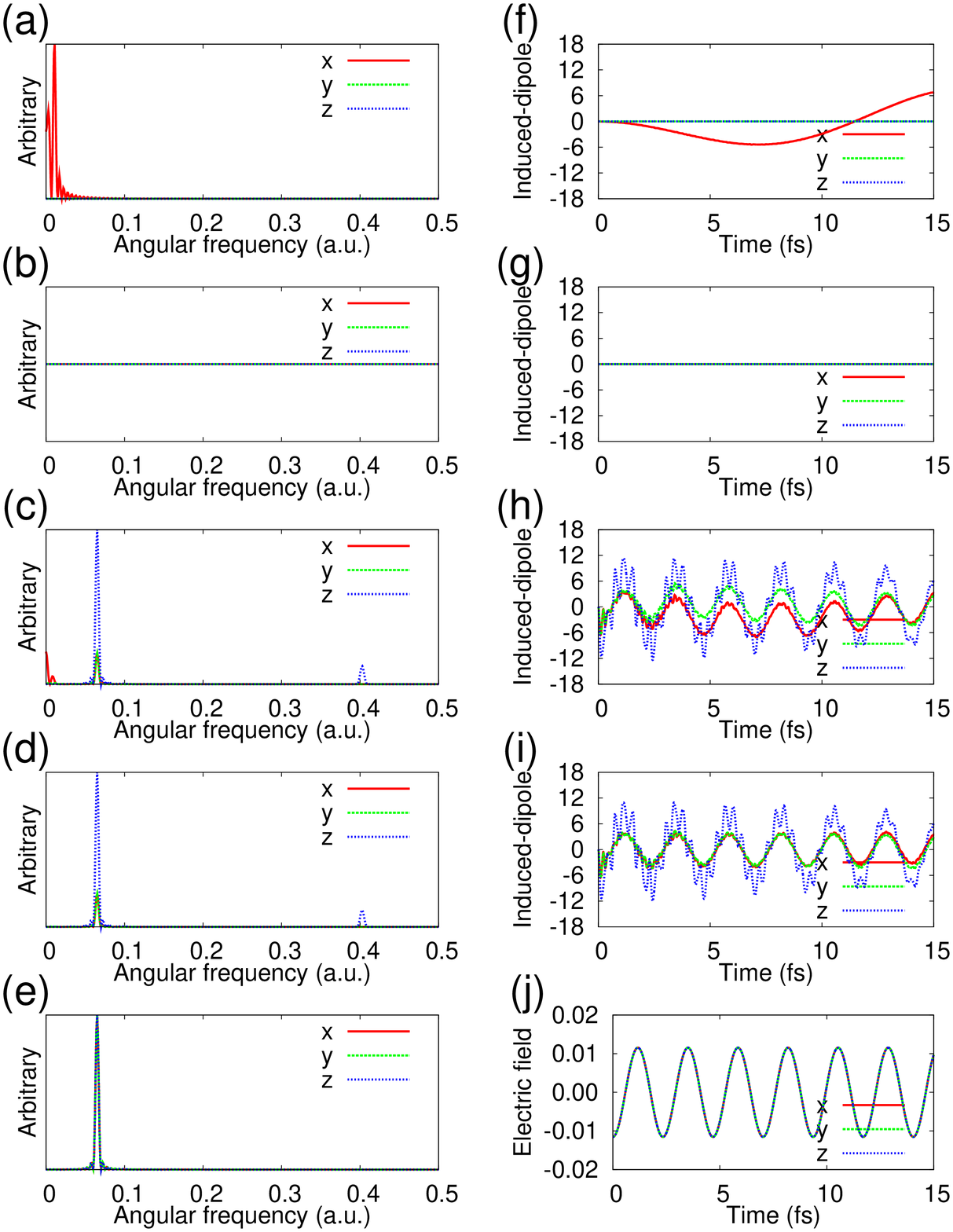}
\caption{Mode analysis of electron 
in the cases of a 20-mer circle of ethylene molecule
with initial excitation. The system is shined by the continuum light 
having 700nm wavelength (corresponding to one photon energy, 0.065 a.u.) 
and normalized polarization vector of (x,y,z) = $(1/\sqrt{3},1/\sqrt{3},1/\sqrt{3})$. 
The panels in left column are the Fourier transformations of 
those placed at just the right position in the figure. 
The panels (a/c) are the Fourier transformations of the time sequences of 
induced dipole moment of the whole system for the cases with 
the initial local excitation of 1st ethylene monomer 
while (b/d) is that for the initial local excitations of 1st and 11th monomers. 
The cases of the panels (a/b) associated correspondingly with (f/g) does not include 
light field while the panels (c/d) corresponding to (h/i) includes. 
(e) is the Fourier transformation of the continuum light field applied in the cases of 
the panels (c/d) and correspondingly (h/i). 
The panels (f/g/h/i) show the induced dipole moment of the whole system 
corresponding to (a/b/c/d) in this order. 
The panel (j) displays the electric field of the light 
in the time domain corresponding to the panel (e).}
\label{FT-RT-ind-dip-Evec-20mer-ethylene}
\end{figure}

Fig. \ref{FT-RT-ind-dip-Evec-20mer-ethylene} gives 
the electronic mode analysis in the cases of a 20-mer circle of ethylene molecule
with initial one and two excitations with and without light field. 
In the case of light, we employed the continuum light 
having 700nm wavelength (corresponding to one photon energy, 0.065 a.u.) 
and normalized polarization vector of (x,y,z) = $(1/\sqrt{3},1/\sqrt{3},1/\sqrt{3})$. 
See also the main article.  
The induced-dipole moment vectors and its Fourier transformation are displayed 
in the same manner as the previous section in this supplementary material.
We find following features which characterize the present dynamics: 
\begin{itemize}
\item [1] Excitation of electronic motion by the light of z polarization is 
          significantly large compared to those of x and y polarization,  
          which is consistent with the non-vanishing z component in the
          transition dipole moment vector with respect to the local HOMO and LUMO of 
          the representative monomer, 1st ethylene, as shown in Tab. \ref{tab-1st-ethylene}. 
\item [2] Electronic motion of z component experience multi photon excitation,  
          which is observed from the spectrum peak around 0.4 a.u. in panels (c) and (d)
          and corresponding high oscillation behavior of z component of the dipole moment 
          in the time domain as found in the panels of (h) and (i). 
          This spectrum peak around 0.4 a.u. is consistent with 
          the local HOMO-LUMO energy difference, 0.0907-(-0.3140) = 0.4047 Hartree, 
          of which values are shown in the caption of Tab. \ref{tab-1st-ethylene}. 
          And also, 700 nm corresponding to the one photon energy 0.065 Hartree 
          is substantially non-resonant to the this typical electron transition. 
\item [3] The case of one local excitation is characterized by the deviation 
          of x and y component as seen in the panels (h). 
          This is reflected in the low frequency spectrum at the position almost near to zero 
          (far lower than 0.065) as seen in the panel (c). 
          The statement that this low frequency mode is attributed to the exciton migration 
          is assisted by the fact that in the light-field free case 
          starting from the initial one site excitation 
          we can find the low frequency mode in the spectrum around zero 
          in the panel (a) and the slow dynamics of the induced dipole moment 
          in the corresponding panel (f). 
\item [4] The low frequency peak seen in the panel (a)  
          does not appear in the case of two local initial excitations 
          as shown in the panel (b). 
          This is cancellation effect due to the symmetry of 
          the two unpaired electron wave packets starting from the 1st and 11th sites 
          at the counter position in the circle. 
          Note that here the dipole moments of 
          total system is analyzed for simplicity. See also the schematic movie 
          of this dynamics, of which explanation is given 
          in the supplemental material. 
          This cancellation effect remains to some extent under the shine of light field 
          as seen in the panels of (d) and (i) though the symmetry is weakly broken as 
          seen in the figure in the main article and the movie. 
\end{itemize}

\section{Suppl.3:Schematics movies of unpaired electron dynamics in 20-mer circle of ethylene molecule}

Here we provide the schematic movies of dynamics of unpaired electron 
in the system of 20-mer ethylene system, 
as mov-20-mer-two-excitons-1.gif and mov-20-mer-two-excitons-2.gif.  
These two files correspond respectively 
to the cases of initial two local site excitations without  
and with light field having the strength being 0.02, wave length of 700nm, 
and polarization vector of 
$
\left(
\dfrac{1}{\sqrt{3}},\dfrac{1}{\sqrt{3}},\dfrac{1}{\sqrt{3}}
\right)
$
. 
The details for plotting data are explained below. 

We constructed the movies by gathering the time sequence of the pictures 
in which the following two dimensional time dependent function, $D(x,y,t)$, 
corresponding to the quantity of unpaired electrons on sites are 
plotted in the x-y plane on which all the center of masses (COMs) of ethylene monomers are placed, 
\begin{align}
D(x,y,t) = \sum_{i=1}^{20} f_i(x,y) \, G_i(t), 
\end{align}
where
\begin{align}
f_i(x,y) = 
\textrm{exp}
\left(  
 - \dfrac{1}{\alpha^2} \left\{  (x-x^{(i)})^2 + (y-y^{(i)})^2  \right\}
\right),
\end{align}
the positions of COM of i-th monomer, 
\begin{align}
x^{(i)} = R \, \textrm{cos}\left( \dfrac{(i-1)\pi}{10} \right) 
\quad  
\textrm{and} 
\quad  
y^{(i)} = R \, \textrm{sin}\left( \dfrac{(i-1)\pi}{10} \right),  
\end{align}
and $G_i(t)$ being the quantity of unpaired electrons in i-th monomer at a time t. 
Here, we set $\alpha = \pi R / 30 $ and $R=12$ (\AA).

\section{Suppl.4:Transition dipole moment vectors and group localized orbital energies of monomers 
in the group diabatic representation}

The transition dipole moment vectors are presented for NPTL and TCNE in the dimer system 
and 1st ethylene in the 20-mer circle of ethylene molecules 
respectively in Tab. \ref{tab-NPTL}, \ref{tab-TCNE} and \ref{tab-1st-ethylene}. 
The associated local orbital energies are also given from the local HOMO-2 to LUMO+2
in the captions of these tables.  
Note that the local properties of monomers presented here 
are the part of the group blocks in the matrix or vectors in the GD representation.
With respect to the HOMO and LUMO energies of monomers, 
we also included in the captions the values obtained 
in the calculations for the totally isolated systems.

%%%%%%%%%%%%%%%%%%%%%%%
%  figures and tables
%%%%%%%%%%%%%%%%%%%%%%%

\clearpage

\begin{table}[t]
\caption{Transition dipole moment vector matrix in the group diabatic representation  
for the NPTL with the distance of molecular planes being 4.0 \AA. Unit is atomic unit. 
Data are shown from 
the local HOMO-2 ( 32-th local orbital state ) to 
    local LUMO+2 ( 37-th local orbital state ) of this monomer. 
Note that the origin employed for the evaluation of the electron position 
is placed at the center of mass of the total dimer system. 
In this order, local MO energies are
$\epsilon^{\textrm{L:HOMO-2}}=-0.3512$ Hartree., 
$\epsilon^{\textrm{L:HOMO-1}}=-0.3012$, 
$\epsilon^{\textrm{L:HOMO}}=-0.2721$, 
$\epsilon^{\textrm{L:LUMO}}=0.0010$,  
$\epsilon^{\textrm{L:LUMO+1}}=0.0316$ and 
$\epsilon^{\textrm{L:LUMO+2}}=0.0792$. 
Here, local HOMO and LUMO are labeled by 34 and 35, respectively.
As a supporting information, 
HOMO and LUMO energies in the totally isolated system are 
-0.2622 Hartree and 0.0115, respectively. 
}%
\label{tab-NPTL}
\begin{center}%
\begin{tabular}{c|ccccccc}
\hline 
\hline 
x & 32  & 33 & 34$^{\textrm{H}}$ & 35$^{\textrm{L}}$ & 36 & 37   \\
\hline
32 &                     -1.0435 &        0.0000 &        2.5248 &        0.0000 &        0.0418 &        0.0000 \\
33 &                      0.0000 &       -1.2461 &        0.0000 &       -2.1075 &        0.0000 &       -0.0478 \\
34$^{\textrm{H}}$ &       2.5248 &        0.0000 &       -1.2746 &        0.0000 &        2.0856 &        0.0000 \\
35$^{\textrm{L}}$ &       0.0000 &       -2.1075 &        0.0000 &       -1.0585 &        0.0000 &       -2.5170 \\
36 &                      0.0418 &        0.0000 &        2.0856 &        0.0000 &       -1.1197 &        0.0000 \\
37 &                      0.0000 &       -0.0478 &        0.0000 &       -2.5170 &        0.0000 &       -1.3105 \\
\hline 
\hline
y & 32  & 33 & 34$^{\textrm{H}}$ & 35$^{\textrm{L}}$ & 36 & 37   \\
\hline
32       &                0.0000 &       -0.0040 &        0.0000 &        0.0736 &        0.0000 &       -1.3900 \\
33       &               -0.0040 &        0.0000 &       -0.0328 &        0.0000 &        1.4397 &        0.0000 \\
34$^{\textrm{H}}$ &       0.0000 &       -0.0328 &        0.0000 &        1.4510 &        0.0000 &        0.0735 \\
35$^{\textrm{L}}$ &       0.0736 &        0.0000 &        1.4510 &        0.0000 &       -0.0231 &        0.0000 \\
36       &                0.0000 &        1.4397 &        0.0000 &       -0.0231 &        0.0000 &        0.0346 \\
37       &               -1.3900 &        0.0000 &        0.0735 &        0.0000 &        0.0346 &        0.0000 \\
\hline 
\hline
z & 32  & 33 & 34$^{\textrm{H}}$ & 35$^{\textrm{L}}$ & 36 & 37   \\
\hline
32 &                     -3.7814 &        0.0000 &       -0.0004 &        0.0000 &        0.0010 &        0.0000 \\
33 &                      0.0000 &       -3.7795 &        0.0000 &        0.0002 &        0.0000 &       -0.0006 \\
34$^{\textrm{H}}$ &      -0.0004 &        0.0000 &       -3.7801 &        0.0000 &       -0.0011 &        0.0000 \\
35$^{\textrm{L}}$ &       0.0000 &        0.0002 &        0.0000 &       -3.7773 &        0.0000 &       -0.0036 \\
36 &                      0.0010 &        0.0000 &       -0.0011 &        0.0000 &       -3.7803 &        0.0000 \\
37 &                      0.0000 &       -0.0006 &        0.0000 &       -0.0036 &        0.0000 &       -3.7759 \\
\hline 
\hline
\end{tabular}
\end{center}%
\end{table}

\begin{table}[t]
\caption{Transition dipole moment vector matrix in the group diabatic representation  
for the TCNE with the distance of molecular planes being 4.0 \AA. Unit is atomic unit. 
Data are shown from 
the local HOMO-2 ( 30-th local orbital state ) to 
    local LUMO+2 ( 35-th local orbital state ) of this monomer. 
Note that the origin employed for the evaluation of the electron position 
is placed at the center of mass of the total dimer system. 
In this order, local MO energies are
$\epsilon^{\textrm{L:HOMO-2}}=-0.4554$ Hartree, 
$\epsilon^{\textrm{L:HOMO-1}}=-0.4420$, 
$\epsilon^{\textrm{L:HOMO}}=-0.3815$, 
$\epsilon^{\textrm{L:LUMO}}=-0.1278$, 
$\epsilon^{\textrm{L:LUMO+1}}=-0.0030$ and 
$\epsilon^{\textrm{L:LUMO+2}}=0.0126$. 
Here, local HOMO and LUMO are labeled by 32 and 33, respectively.
As a supporting information, 
HOMO and LUMO energies in the totally isolated system are 
-0.3901 Hartree and -0.1367,respectively. 
}%
\label{tab-TCNE}
\begin{center}%
\begin{tabular}{c|ccccccc}
\hline 
\hline 
x & 30  & 31 & 32$^{\textrm{H}}$ & 33$^{\textrm{L}}$ & 34 & 35   \\
\hline
30   &   1.0762   &   0.0263  &    0.0000 &    1.6083  &    0.0000 &    -0.0036 \\
31   &   0.0263   &   1.0362  &    0.0000 &    -0.0052  &    0.0000 &    -1.3513 \\
32$^{\textrm{H}}$   &   0.0000   &   0.0000  &    1.1384 &    0.0000  &   -0.0025 &    0.0000 \\
33$^{\textrm{L}}$   &   1.6083   &  -0.0052  &    0.0000 &    1.1767  &    0.0000 &    0.0035 \\
34   &   0.0000   &   0.0000  &   -0.0025 &    0.0000  &    1.2254 &    0.0000 \\
35   &  -0.0036   &  -1.3513  &    0.0000 &    0.0035  &    0.0000 &    1.1979 \\
\hline 
\hline
y & 30  & 31 & 32$^{\textrm{H}}$ & 33$^{\textrm{L}}$ & 34 & 35   \\
\hline
30   &   0.0000   &   0.0000   &   0.0546 &    0.0000  &   -0.0033 &    0.0000 \\
31   &   0.0000   &   0.0000   &  -0.0048 &    0.0000  &    0.0107 &    0.0000 \\
32$^{\textrm{H}}$   &   0.0546   &  -0.0048   &   0.0000 &    -2.1112  &    0.0000 &    -0.0039 \\
33$^{\textrm{L}}$   &   0.0000   &   0.0000   &  -2.1112 &    0.0000  &   -0.0104 &    0.0000 \\
34   &  -0.0033   &   0.0107   &   0.0000 &    -0.0104  &    0.0000 &    2.7252 \\
35   &   0.0000   &   0.0000   &  -0.0039 &    0.0000  &    2.7252 &    0.0000 \\
\hline 
\hline
z & 30  & 31 & 32$^{\textrm{H}}$ & 33$^{\textrm{L}}$ & 34 & 35   \\
\hline
30   &   3.7800  &   -0.0504  &    0.0000 &    0.0004 &    0.0000 &    -0.0006 \\
31   &  -0.0504  &    3.7813  &    0.0000 &    0.0005 &    0.0000 &    0.0000 \\
32$^{\textrm{H}}$   &   0.0000  &    0.0000  &    3.7827 &    0.0000 &    -0.1305 &    0.0000 \\
33$^{\textrm{L}}$   &   0.0004  &    0.0005  &    0.0000 &    3.7868 &    0.0000 &    -0.0451 \\
34   &   0.0000  &    0.0000  &   -0.1305 &    0.0000 &    3.7833 &    0.0000 \\
35   &  -0.0006  &    0.0000  &    0.0000 &    -0.0451 &    0.0000 &    3.7815 \\
\hline 
\hline
\end{tabular}
\end{center}%
\end{table}

\begin{table}[t]
\caption{Transition dipole moment vector matrix in the group diabatic representation  
for the 1st ethylene monomer in the 20-mer ethylene aggregate system. Unit is atomic unit. 
Data are shown from 
the local HOMO-2 ( 6-th local orbital state ) to 
    local LUMO+2 ( 11-th local orbital state ) of this monomer. 
Note that the origin employed for the evaluation of the electron position 
is placed at the center of mass of the total dimer system. 
In this order, local MO energies are 
$\epsilon^{\textrm{L:HOMO-2}}=-0.4732$ Hartree, 
$\epsilon^{\textrm{L:HOMO-1}}=-0.4056$, 
$\epsilon^{\textrm{L:HOMO}}=-0.3140$, 
$\epsilon^{\textrm{L:LUMO}}=0.0907$, 
$\epsilon^{\textrm{L:LUMO+1}}=0.1819$ and 
$\epsilon^{\textrm{L:LUMO+2}}=0.2014$.
Here, local HOMO and LUMO are labeled by 8 and 9, respectively.
As a supporting information, 
HOMO and LUMO energies in the totally isolated system are 
-0.3277 Hartree and 0.0756,respectively. 
}%
\label{tab-1st-ethylene}
\begin{center}%
\begin{tabular}{c|ccccccc}
\hline 
\hline 
x & 6  & 7 & 8$^{\textrm{H}}$ & 9$^{\textrm{L}}$ & 10 & 11   \\
\hline
6 &   22.6769&     0.0000&     0.0000&     0.0000&    -0.0211&    -0.8328 \\
7 &    0.0000&    22.6762&     0.0000&     0.0000&     0.0000&     0.0000 \\
8$^{\textrm{H}}$ &    0.0000&     0.0000&    22.6798&     0.0000&     0.0000&     0.0000 \\
9$^{\textrm{L}}$ &    0.0000&     0.0000&     0.0000&    22.6812&     0.0000&     0.0000 \\
10 &   -0.0211&     0.0000&     0.0000&     0.0000&    22.7776&     1.8637 \\
11 &   -0.8328&     0.0000&     0.0000&     0.0000&     1.8637&    22.5812 \\
\hline 
\hline
y & 6  & 7 & 8$^{\textrm{H}}$ & 9$^{\textrm{L}}$ & 10 & 11   \\
\hline
6 &    0.0000&     0.0000&     0.0430&     0.0000&     0.0000&     0.0000 \\
7 &    0.0000&     0.0000&     0.0000&    -0.0031&     0.0000&     0.0000 \\
8$^{\textrm{H}}$ &    0.0430&     0.0000&     0.0000&     0.0000&     0.0895&    -0.0058 \\
9$^{\textrm{L}}$ &    0.0000&    -0.0031&     0.0000&     0.0000&     0.0000&     0.0000 \\
10 &    0.0000&     0.0000&     0.0895&     0.0000&     0.0000&     0.0000 \\
11 &    0.0000&     0.0000&    -0.0058&     0.0000&     0.0000&     0.0000 \\
\hline 
\hline
z & 6  & 7 & 8$^{\textrm{H}}$ & 9$^{\textrm{L}}$ & 10 & 11   \\
\hline
6 &    0.0000&    -0.0003&     0.0000&     0.0000&     0.0000&     0.0000 \\
7 &   -0.0003&     0.0000&     0.0000&     0.0000&    -0.0254&    -1.0482 \\
8$^{\textrm{H}}$ &    0.0000&     0.0000&     0.0000&    -1.3581&     0.0000&     0.0000 \\
9$^{\textrm{L}}$ &    0.0000&     0.0000&    -1.3581&     0.0000&     0.0000&     0.0000 \\
10 &    0.0000&    -0.0254&     0.0000&     0.0000&     0.0000&     0.0000 \\
11 &    0.0000&    -1.0482&     0.0000&     0.0000&     0.0000&     0.0000 \\
\hline 
\hline
\end{tabular}
\end{center}%
\end{table}

% ===============
%\begin{thebibliography}{99}                                                   
%
%%
%\bibitem{S-omega}
%U. N. Morzan, F. F. Ramírez, M. B. Oviedo, C. G. S\'anchez, D. A. Scherlis, and M. C. G. Lebrero
%J. Chem. Phys. \textbf{140}, 164105 (2014). 
%
%\end{thebibliography}

\end{document}